\documentclass[12pt,psfig,a4]{article}
\usepackage{geometry}
\usepackage{amsmath}
\usepackage{graphics}
\usepackage{graphicx,color}
\usepackage{setspace}
\usepackage{mathrsfs}
\usepackage{amsbsy,amssymb,amsmath,ulem}
\usepackage{hyperref}
\hypersetup{colorlinks=true,
			linkcolor=magenta,
			citecolor=blue}

\newcommand{\scE}{\mathscr{E}}
\newcommand{\scU}{\mathscr{U}}

\newcommand{\PHI}{{\boldsymbol{\Phi}}}

\newcommand{\DD}{{\boldsymbol{D}}}

\newcommand{\EE}{{\boldsymbol{E}}}
\newcommand{\FF}{{\boldsymbol{F}}}
\newcommand{\AAA}{{\boldsymbol{A}}}

\newcommand{\JJ}{{\boldsymbol{J}}}

\newcommand{\ww}{{\boldsymbol{w}}}
\newcommand{\xx}{{\boldsymbol{x}}}
\newcommand{\yy}{{\boldsymbol{y}}}
\newcommand{\ff}{{\boldsymbol{f}}}
\newcommand{\zz}{{\boldsymbol{z}}}

\newcommand{\HH}{{\boldsymbol{H}}}
\newcommand{\BB}{{\boldsymbol{B}}}
\newcommand{\GG}{{\boldsymbol{G}}}
\newcommand{\rr}{{\boldsymbol{r}}}

\newcommand{\qq}{{\boldsymbol{q}}}
\newcommand{\cc}{{\boldsymbol{c}}}
\newcommand{\QQ}{{\boldsymbol{Q}}}
\newcommand{\PP}{{\boldsymbol{P}}}
\newcommand{\II}{{\boldsymbol{I}}}
\newcommand{\TT}{{\boldsymbol{T}}}

\newcommand{\pp}{{\boldsymbol{p}}}
\newcommand{\vv}{{\boldsymbol{v}}}

\newcommand{\nn}{{\boldsymbol{n}}}

\newcommand{\Lyx}{L\kern-.1667em\lower.25em\hbox{y}\kern-.125emX\spacefactor1000}
\begin{document}
\pagestyle{plain}
\pagenumbering{arabic}


\title{\bf \large Solid-fluid dynamics of yield-stress fluids}
\author{
	Ilya Peshkov\thanks{Ecole Polytechnique de Montreal, Canada. On leave from Sobolev Institute of Mathematics, Russian Academy of Sciences, Russia; \url{peshkov@math.nsc.ru}},\ \ \ 
	Miroslav Grmela\thanks{Ecole Polytechnique de Montreal, Canada; \url{miroslav.grmela@polymtl.ca}},\ \ \
	Evgeniy Romenski\thanks{Sobolev Institute of Mathematics, Russian Academy of Sciences, Russia; \url{evrom@math.nsc.ru}}
	\vspace{20pt}
	}

\date{}
\maketitle


\begin{abstract}

On the example of two-phase continua experiencing stress induced solid-fluid phase transitions we explore the use of the Euler structure in the formulation of the governing equations. The Euler structure guarantees that solutions of the time evolution equations possessing it are compatible with mechanics and with thermodynamics. The former compatibility means that the equations are local conservation laws of the Godunov type and the latter compatibility means  that the entropy does not decrease during the time evolution. In numerical illustrations,  in which the one-dimensional Riemann problem is explored,  we require that the Euler structure is also preserved in the discretization.

\end{abstract}


\tableofcontents

\section{\label{Intro}Introduction}

The strategy  that we use in the formulation of  equations governing the time evolution of a macroscopic system is based on the requirement that their solutions (i.e.  predictions of the model) agree with certain basic experimental observations. These observations are of two types: mechanical and thermodynamical. The mechanical observations consist of observations of some fundamental consequences of Newton's law (as  for example  conservations of  the overall mass, energy, and momentum). The thermodynamic observations are observations constituting the experimental basis of classical equilibrium thermodynamics (i.e. observations of the approach to thermodynamic equilibrium at which the behavior is found to be well described by the classical equilibrium thermodynamics).

What are the mathematical structures guaranteeing the compatibility with mechanics and thermodynamics?
There are two such structures. The first one originates in Euler's formulation of
Newton's law   in the setting of continuum mechanic \cite{Euler}. Its extension  to thermodynamics has been developed in nonequilibrium thermodynamics (see e.g. Ref.~\cite{deGM}) and later in Refs.~\cite{god1961,lax1971,god1972,fried1978,god_rom2003,rugg2005} where also   mathematical aspects of the formulation have  been addressed.  We shall refer to the  structure that unfolded from \cite{Euler} as the Euler structure.  The second structure addressing the compatibility with mechanics and thermodynamics has unfolded from Clebsch's reformulation of Euler's equations into the Hamiltonian form \cite{Clebsch}. Its thermodynamics extension has been proposed in Refs.~\cite{G1,G2,G22,G3,G4,BE,G5,G6,G7,G8,G9}. We shall refer to the second structure as the Clebsch structure. Since continuum mechanics can be formulated in two types of coordinates, namely in the Euler and the Lagrange coordinates, we shall use the terms Euler structure in the Lagrangian framework (discussed in Section~\ref{LF}) and Euler structure in the Eulerian framework (discussed in Section~\ref{sec:eul_dis}).

Both the Euler and the Clebsch structures have their advantages and disadvantages. The advantage of the Euler structure is that it addresses also  mathematical regularity of the formulation. Its main disadvantage is its limitation to formulations involving only a certain type of partial differential equations (namely hyperbolic partial differential equations expressing local conservations). On the other hand,  the main advantage of the Clebsch structure is its universal applicability (due to the universal applicability of the concept of the abstract Hamiltonian structure) and its direct thermodynamic interpretation (the time evolution can be seen as a continuous sequence of Legendre transformations maximizing the entropy -- see Ref.~\cite{G8}).

Our objective in this paper is to compare the Clebsch and the Euler structures and then to explore the use of the Euler structure in the context of two-phase solid-fluid continua experiencing stress induced solid-fluid transitions. Such continua are for example yield-stress fluids, also called viscoplastic fluids,  (see e.g. \cite{putz,Frigaard2014})  that behave like solids in  unyielded regions (i.e. below a critical applied stress called a yield stress) and like liquids in yielded regions (i.e. at higher  stresses). In this paper we do not aim at formulating a complete model of yield-stress fluids that would also include  a  comparison of  its predictions with results of experimental observations. We are making only preparatory investigations in this direction.

Having the mathematical  structure, the governing equations of the model are formulated as its particular realization. By this we mean that all the abstract elements involved in the structure acquire a concrete form (as for instance abstract elements of a group become matrices in group representations). It is in the realizations (presented in
Section \ref{LF} in the Lagrangian  framework and in Section  \ref{sec:eul_dis} in the Eulerian  framework) where  the physical insight into the specific nature of the system under consideration enters the construction of governing equations. We then also use the Euler structure in numerical calculations in which we illustrate certain aspects of solutions of the governing equations.

\section{\label{Euler_struct}Euler Structure}

Let
\begin{equation}\label{x}
\boldsymbol{q}(\rr)\in\mathbb{R}^n
\end{equation}
be a set of fields playing the role of state variables. By $\rr=(r_1,r_2,r_3)^\mathsf{T}\in\Omega\subset\mathbb{R}^3$ we denote the position vector. The only structure that we impose on  $\boldsymbol{q}(\rr)$ is the fibration
\begin{equation}\label{ez}
\boldsymbol{q}(\rr)=(e(\rr),\boldsymbol{q}'(\rr))
\end{equation}
where $e(\rr)$, called energy, is a scalar field and $\boldsymbol{q}'(\rr)$ are the remaining fields.

The vector field generating the time evolution of  $\boldsymbol{q}(\rr)$ is a sum of two parts: one, called nondissipative, is denoted $\left(\frac{\partial}{\partial t}\right)_{nondiss}$ and the other, called dissipative, is denoted by  $\left(\frac{\partial}{\partial t}\right)_{diss}$, i.e.
\begin{equation}\label{nd}
\frac{\partial}{\partial t}=\left(\frac{\partial}{\partial t}\right)_{nondiss} + \left(\frac{\partial}{\partial t}\right)_{diss}
\end{equation}
We now list requirements put on the nondissipative and the dissipative vector fields.

\subsection{\label{nvf}Nondissipative vector field}

The nondissipative vector field is given by
\begin{equation}\label{xt}
\left(\frac{\partial \boldsymbol{q}}{\partial t}\right)_{nondiss}+{\rm div}\boldsymbol{\mathcal{F}}(\qq )=0
\end{equation}
By the symbol   $\boldsymbol{\mathcal{F}}=[\mathcal{F}_i^j]$, $i=1,2,\ldots,n$, $j=1,2,3$ we denote   fluxes corresponding to the fields  (\ref{ez}). They are themselves fields depending on
$\boldsymbol{q}(\rr)$.
The time evolution equation (\ref{xt}) is called a local conservation law since it implies $\frac{{\rm d}\boldsymbol{Q}}{{\rm d}t}=0$, where  $\boldsymbol{Q}=\int_{\Omega} d\rr\boldsymbol{q}(\rr)$,   provided appropriate boundary conditions have been chosen (i.e. boundary conditions guaranteeing $\int_{\Sigma} \boldsymbol{\mathcal{F}}\cdot\nn = 0$; $\Sigma$ denotes the boundary of $\Omega\subset\mathbb{R}^3$ and $\nn$ the vector perpendicular to the boundary). In particular, due to the presence of the energy field in the set of state variables (see (\ref{ez})), the total energy $E=\int d\rr e(\rr)$ is conserved.

We shall make now some requirements on the fluxes $\boldsymbol{\mathcal{F}}$.

The first requirement has purely mathematical origins. We require that $\boldsymbol{\mathcal{F}}$ depends on the fields $\boldsymbol{q}(\rr)$ but not on their derivatives. This requirement
enormously simplifies the mathematical considerations (as e.g. the question of well posedness of the initial value problem and the numerical analysis) while not constraining (after making appropriate modifications and changes) the physical content of the analysis.

The second requirement is of the physical origin but, as we shall see, with very important mathematical consequences. We require that (\ref{xt}) implies another local conservation law
\begin{equation}\label{So}
\left(\frac{\partial s(\boldsymbol{q} )}{\partial t}\right)_{nondiss}+{\rm div} \boldsymbol{F}^{(s)}(\boldsymbol{q} )=0
\end{equation}
where  $s(\boldsymbol{q} )$ and  $ \boldsymbol{F}^{(s)}=\left(F^{(s)}_1,F^{(s)}_2,F^{(s)}_3\right)^\mathsf{T}$, called entropy field and entropy flux respectively,  are functions of $\boldsymbol{q}(\rr)$.  This additional conservation law is the reason why the vector field $\left(\frac{\partial}{\partial t}\right)_{nondiss}$  is called a nondissipative vector field.  As for the   functional dependence of $s$ on $\boldsymbol{q}$, we require that
\begin{equation}\label{sprop}
\begin{array}{l}
-s(\qq)\ \ \mbox{is a convex function of }\qq,\\[3mm]
\displaystyle\frac{\partial s}{\partial e}>0.
\end{array}
\end{equation}The first requirement expresses the thermodynamic stability and the second the positivity of the absolute temperature (recall that $\frac{\partial s}{\partial e}=\frac{1}{T}$, where $T$ is the local  absolute temperature).

Godunov has noted \cite{god1961}  that    for a subclass of local conservation laws  (\ref{xt})
(called hereafter a Godunov class of conservation laws) that can be written in the form
\begin{equation}\label{god_form1}
\left(\frac{\partial  L_\pp}{\partial  t}\right)_{nondiss}+\frac{\partial M^j_\pp}{\partial r_j}=0
\end{equation}
an additional  conservation law
\begin{equation}\label{god_energy}
\left(\frac{\partial (p_k L_{p_k}-L)}{\partial t}\right)_{nondiss} +\frac{\partial (p_k M^j_{p_k}-M^j)}{\partial r_j}=0
\end{equation}
is automatically implied (it is sufficient to multiply each equation in (\ref{god_form1}) by $p_k$ and sum them up). Here,
$\boldsymbol{p}(\rr)=(p_1,p_2,\ldots,p_n)^\mathsf{T}$ is a vector of state variables,
$L(\pp)$ and $M^j(\pp)$ are potentials ($L(\pp)$ is in addition required to be convex). Moreover, (\ref{god_energy}) implies also
\begin{equation}\label{eq:god_sym}
L_{\pp\pp}\left(\frac{\partial \pp}{\partial t}\right)_{nondiss}+M^j_{\pp\pp}\frac{\partial \pp}{\partial r_j}=0
\end{equation}
which means that (\ref{god_energy}) is a symmetric hyperbolic (or hyperbolic in the sense  of Friedrichs \cite{fried1958}) system of partial differential equations. This in turn  means that the initial value problem (Cauchy problem) for  (\ref{god_form1}) with sufficiently smooth initial data is well posed. Such connection between thermodynamics and well-posedness of differential equations of continuum mechanics was first recognized by Godunov in \cite{god1961}  (see also Refs.~\cite{lax1971,god1972,god_rom2003,god1995,god1996,rom1998,rom2001}). The notation that we have used in the equations above and that we shall continue to use it in the rest of this paper is the following: $ L_{\pp}=\partial L/\partial \pp=(L_{p_1},L_{p_1},\ldots,L_{p_n})^\mathsf{T}$ is $n$-vector, $L_{\pp\pp}=\partial^2 L /\partial \pp^2=[L_{p_ip_j}]$ is $n\times n$-matrix, and similarly for other potentials; moreover,  we use also the summation convention (i.e. summation over  repeated indices).

Equation   (\ref{god_form1}) is related to (\ref{xt}) by Legendre transformation. Indeed, we see clearly that
\begin{equation}\label{eq:p_q}
\qq=L_\pp, \ \ \ \pp=s_{\qq},
\end{equation} and entropy $s(\qq)$ is the Legendre transformation of $L(\pp)$, i.e. $s(\qq)=p_k L_{p_k}-L(\pp)$, and $F^{(s)}_j(\qq)=p_k M^j_{p_k}-M^j(\pp)$. The Legendre transformation of (\ref{god_form1}) defines the Godunov class inside the class (\ref{xt}) of  local conservation laws.

\subsection{\label{dvf}Dissipative vector field}

The dissipative vector field $\left(\frac{\partial}{\partial t}\right)_{diss}$ generates the time evolution during which the energy is still conserved but the entropy $s(\boldsymbol{q})$, introduced in (\ref{So}),  increases. It is this  property of entropy generation that gives the vector field $\left(\frac{\partial}{\partial t}\right)_{diss}$ its name. The complete vector field (\ref{god_form1}) is thus (hereafter we consider only the Godunov class (\ref{god_form1}) of the conservation laws)
\begin{equation}\label{xtcomp}
\frac{\partial  L_\pp}{\partial  t}+\frac{\partial M^j_\pp}{\partial r_j}=\boldsymbol{\mathcal{S}}(\pp )
\end{equation}
where $\boldsymbol{\mathcal{S}}=(\mathcal{S}_1,\mathcal{S}_2,\ldots,\mathcal{S}_n)^\mathsf{T}$ is the dissipative vector field (we shall refer to it as source terms). The general requirements restricting the choice of $\boldsymbol{\mathcal{S}}$ are: \texttt{(i)} energy remains conserved, \texttt{(ii)} entropy does not decrease, and \texttt{(iii)} $\boldsymbol{\mathcal{S}}$
is a function of $\boldsymbol{p}(\rr)$ but not of derivatives of $\boldsymbol{p}(\rr)$ with respect to $\rr$. We shall see examples of the source terms satisfying these three requirements below in the following sections. In Comment 3 below in this section we explain that the requirement \texttt{(iii)}  does not exclude for example the very frequently used Fourier and the Navier-Stokes dissipation.

With the time evolution governed by the non-homogeneous conservation laws (\ref{xtcomp}) the
local conservation law (\ref{So}) is replaced by another  non-homogeneous conservation law
\begin{equation}\label{Sod}
\frac{\partial (p_k L_{p_k}-L)}{\partial t} + \frac{\partial (p_k M^j_{p_k}-M^j)}{\partial r_j}=\varsigma(\boldsymbol{p})
\end{equation}
where $\varsigma(\boldsymbol{q})=-\pp^\mathsf{T}\boldsymbol{\mathcal{S}}(\qq)=-p_k\mathcal{S}_k$ is the  entropy production.  The thermodynamic compatibility requires that (\ref{xtcomp}) remains symmetric hyperbolic, that the energy field $e(\rr)$ (see (\ref{ez})) remains to be a local conservation law  (this  means that the total energy $E=\int_{\Omega}d\rr e(\rr)$ is   conserved even in the presence of dissipation), and that entropy production $\varsigma(\boldsymbol{q} )$ is non negative (i.e. $\varsigma(\boldsymbol{q} )\geq 0$).

All the time evolution equations arising in this paper will be  cast (or at least  attempted to be cast) into the form of the Godunov equations (\ref{god_form1}).

\subsection{\label{comments}Comments}

We end this section with several  comments.


\subsubsection*{Comment 1}

The second property in (\ref{sprop}) makes it possible to exchange the energy field $e(\rr)$ in the set of state variables with the entropy field $s(\rr)$. Using the terminology of Callen \cite{Callen}, the choice of state variables made in (\ref{ez}) represents \textit{the entropy representation} and the choice of the fields
\begin{equation}\label{sz}
\boldsymbol{q}(\rr)=(s(\rr),\boldsymbol{q}'(\rr))
\end{equation}
as state variables represents \textit{the energy representation}. In the investigation of particular realizations of the Euler structure, it turns out to be convenient and natural to use the energy representation in discussions of the nondissipative vector field (in this representation it is easier to guarantee  the requirement (\ref{So})) and  the entropy representation in discussions of the dissipative vector field (in this representation it is easier to guarantee  the requirement of the energy conservation).  In this paper we shall however use the energy representation  (\ref{sz}) in  both dissipative and nondissipative dynamics.
\\

\subsubsection*{Comment 2}

A particular realization of the Euler structure consists of the following four steps:
\begin{itemize}

\item
Specification of the fields $\boldsymbol{q}(\rr)$,

\item
Specification of the fluxes $\boldsymbol{\mathcal{F}}(\boldsymbol{q} )$ (in the setting of classical fluid mechanics, this specification is  called  a constitutive relation),

\item
Specification of the entropy $s(e,\boldsymbol{q}' )$ (or alternatively  $e(s,\boldsymbol{q}' )$) called a fundamental thermodynamic relation,

\item
Specification of the entropy flux $\boldsymbol{F}^{(s)}(\boldsymbol{q} )$ (or alternatively $\boldsymbol{F}^{(e)}(\boldsymbol{q} )$ the energy flux).
\end{itemize}

We shall follow these four steps in Section \ref{LF}.
\\

\subsubsection*{Comment 3}

The third comment is about the limitation to hyperbolic  partial differential equations. The limitation  seems to be  severe since it excludes very frequently used fluid models like for instance the Fourier model of heat conduction  or the Navier-Stokes model of viscous fluids. It has been however realized \cite{Catt,EIT}  that both of these models as well as many other models  of the same type can be lifted to larger spaces (by adopting extra state variables) in which the time evolution equations are hyperbolic.
For example, the Fourier theory, if lifted to a larger space involving  the heat flux (or a related to it field) as an extra state variable, becomes a Cattaneo \cite{Catt} theory in which the governing equations form  a system of hyperbolic partial differential equations. Inside the Cattaneo  theory, the original Fourier theory appears  in the limit when  one of the parameters introduced in  the extended theory (namely the relaxation time of the heat flux) tends to zero and the heat flux ceases to be an independent state variable (it becomes enslaved to the fields forming the set  of state variables in the Fourier theory). From the physical point of view, the lift from the Fourier to the Cattaneo theory can be interpreted as an inclusion of inertia into the time evolution of heat. Similarly, the Navier-Stokes fluid equations can be lifted to a hyperbolic system of partial differential equations by adopting the stress tensor (or related to it field) as an extra state variable \cite{EIT}. Alternatively, the extra field can also be chosen in such a way that they characterize the motion on the microscopic scale. If we recall that the physical origin of the dissipation is considered to lie  in the microscopic motion, we see that an explicit introduction of dissipation into the microscopic motion and leaving the equations governing the classical macroscopic fields without an explicit dissipation is physically meaningful. The macroscopic fields will dissipate indirectly through their coupling with the microscopic fields.
\\

\subsubsection*{Comment 4}

The  Clebsch structure \cite{G1,G2,G22,G3,G4,BE,G5,G6,G7,G8,G9}  differs from the Euler structure in the following points:
\begin{itemize}

\item[(C1)] The state variables $\qq$ that are admissible in the analysis that use the Clebsch structure are not restricted to  fields $\boldsymbol{q}(\rr)$. They   can be distribution functions (as it is the case in kinetic theories) or finite dimensional vectors (as it is the case for example in complete microscopic theories in which macroscopic systems are seen as composed of a finite number of particles, or in chemical kinetics where $\boldsymbol{q}$ is a vector whose components are number of moles of a finite number of components).

\item[(C2)]  The time evolution equation (\ref{xt}) is  replaced in the Clebsch structure by Hamilton's  equation $\left(\frac{\partial \boldsymbol{q}}{\partial t}\right)_{nondiss}=L(\boldsymbol{q}) E_{\boldsymbol{q}}$, where $L(\boldsymbol{q})$ is a Poisson bivector (i.e. $<a_{\boldsymbol{q}},L(\boldsymbol{q})b_{\boldsymbol{q}}> $ is a Poisson bracket that we denote $\{a,b\}$;  $a$ and $b$ are real valued functions of $\boldsymbol{q}$ and $<,>$ denotes a scalar product).

\item[(C3)]  The local conservation law (\ref{So}) is replaced by a global conservation $\left(\frac{\partial S}{\partial t}\right)_{nondiss}=0$, where $S=\int_{\Omega} d\rr s(\boldsymbol{q} )$. In view of $\left(\frac{\partial \boldsymbol{q}}{\partial t}\right)_{nondiss}=L(\boldsymbol{q}) E_{\boldsymbol{q}}$,  this requirement takes the form $\{a,S\}=0$ for all functions $a$, or in other words, the Poisson bracket $\{a,b\}$ is required to be degenerate and the entropy $S$ is its Casimir function.

\item[(C4)] The requirement of the energy local conservation law $\frac{\partial e}{\partial t}=-{\rm div}\FF^{(e)}$, where $\FF^{(e)}$ is the energy flux, is replaced by the requirement of the global energy conservation $\frac{dE}{dt}=0$.
\end{itemize}

\subsubsection*{Comment 5}

What is common to the Euler and the Clebsch structures (see in particular  the formulation of the Clebsch structure developed in Ref.~\cite{G7,G8})  is a systematic  use of both the state variables $\boldsymbol{q}$ and their conjugates $\pp$,  and of both the generating potentials and their Legendre transformations. In the Clebsch structure this duality is then manifestly displayed and used by  placing the time evolution into the setting of contact geometry in which the Legendre transformations are the natural transformations (similarly as the rotations are natural transformations in the metric geometry).

We hope to bring more light into the relationship between the Euler and the Clebsch structures in a future paper.

\section{\label{LF}Euler Structure in the Lagrangian Framework: Particular Realization for Solid-Fluid Mixture}

We proceed now to construct a particular realization of the Euler structure expressing the behavior of solid-fluid mixtures experiencing irreversible deformation through the stress driven phase transition. First,  we discuss the Euler structure within the Lagrangian framework (Lagrangian model) and then we complete the realization with the analysis of the Euler structure in the Eulerian framework (Eulerian model). We  shall follow the steps listed in the second comment in Section \ref{comments}.

\subsection{State variables }

We begin our investigation with the fields
\begin{equation}\label{stvnd}
\boldsymbol{q}=(\vv,\FF,c,\hat{\ww},\alpha,S_1,S_2,\boldsymbol{P})
\end{equation}
serving as state variables. Their physical meaning is the following. The  scalar field  $c$ is the mass fraction of the first component (i.e. the solid phase), $\alpha$ is another scalar field denoting the volume fraction of the first component. If $\rho_1$ and $\rho_2$ are the mass densities of the first and the second components then the total mass density $\rho=\alpha \rho_1+(1-\alpha)\rho_2$  and $c=\frac{\alpha \rho_1}{\rho}$. The fields $\rho,c,\alpha$ are considered to be mutually independent due to the  heterogeneity  of the solid-fluid mixture. A more detailed consideration of its morphology (that may include for instance characterization of  shapes of  solid inclusions) is outside the mesoscopic viewpoint taken in this paper. 

Let $s$ be the overall entropy field appearing in (\ref{So}) and (\ref{Sod}). If we denote by $s_1$ and $s_2$ the entropies of the first and the second components respectively then $s=cs_1+(1-c)s_2\equiv S_1+S_2$, where
the fields $S_1=cs_1$, $S_2=(1-c)s_2$ are the partial entropies of the components. By considering  two separate entropies $s_1$, $s_2$ as independent state variables we allow the two components to have different temperatures (recall that the temperature of the component $i$ is defined by $T_i=\frac{\partial e_i}{\partial s_i}$ if $e_i$ denotes the $i$-th phase internal energy).

Let $\vv_1$ and $\vv_2$ denote velocities of the first and the second component respectively. Then $\vv=c\vv_1+(1-c)\vv_2$ and $\ww=\vv_1-\vv_2$ be the mixture velocity and the relative velocity respectively.

The tensor field $\FF$ is the overall deformation gradient tensor: $\FF=[F_{ij}]=\frac{\partial \xx}{\partial \yy}=\left[\frac{\partial x_i}{\partial y_j}\right]$, where $\xx=(x_1,x_2,x_3)^\mathsf{T}$ denote the coordinates of the position vector, $\rr$, at the  time $t$ relative to a Cartesian coordinate system, $\yy=(y_1,y_2,y_3)^\mathsf{T}$ denote the coordinates of $\rr$ at the initial moment of time, $t=0$, relative to the same coordinate system. The coordinates $\xx$ are also called the Eulerian coordinates and $\yy$ are called label (Lagrangian) coordinates. These two type of coordinates are related by the following system of ordinary differential equations:
\begin{equation}\label{Lag}
\frac{{\rm d}\xx(t,\yy)}{{\rm d} t}=\vv(\xx,t);\,\,\,\,\xx(0,\yy)=\yy.
\end{equation}

The strain $\FF$ and the mixture mass density $\rho$ are related by the equality:
\[\det\FF=\rho_0/\rho,\]
where $\rho_0$ is the reference mass density of the mixture, i.e. $\rho_0=\rho$ when $\yy=\xx$.

We  denote  the Lagrangian relative velocity by the symbol $\hat{\ww}=(\hat{w}_1,\hat{w}_2,\hat{w}_3)^{\mathsf{T}}= \FF^{\mathsf{T}}\ww$. It turns out that the relative velocity $\ww$ is not a conserved quantity in the Lagrangian framework. That is the reason why, in this section, we consider $\hat{\ww}$ as a state variables instead of $\ww$.  Conversely, in the Eulerian frame,  we will use  $\ww$ as the state variable (see (\ref{eul_q})).

The tensor $\PP=[P_{ij}]$ describes irreversible (plastic) deformations.
In addition to the two strain tensors $\FF$ and $\PP$, we also introduce another tensor $\EE$ that is related to $\FF$ and $\PP$  by
\begin{equation}\label{FEP}
\FF=\EE\PP
\end{equation}
The tensor $\EE$ describes elastic deformations. We discuss the physical interpretation of all three tensors $\FF$,  $\PP$, and $\EE$ in more detail in Section \ref{sec: Maxwell}. Here, we only mention that the tensor $\PP$ is constrained by requiring
\begin{equation}\label{detP}
\det\PP=1,
\end{equation}
i.e. $\det\FF=\det\EE$. From the physical point of view, this constraint means that we consider only  inelastic deformations which do not change the mass of a control volume (see more in Section \ref{sec: Maxwell}).

We have chosen the state variables (\ref{stvnd})  because  they  appear to us as  most appropriate. Another possible choice would be, for example,  to replace
the deformation gradient $\FF$ with the displacement field. The time evolution equations with such state variables cannot be however cast into the Godunov form. On the other hand, as we shall see below,  equations governing the time evolution of  (\ref{stvnd}) do possess the Godunov structure and we can thus benefit from all general implications  (concerning the physical and the mathematical regularity) of the structure.   Still another choice could be to replace the deformation gradient $\FF$ with two deformation gradients, each addressing deformations of the individual phases.
Such choice
would however lead to two  different  Lagrangian coordinates and consequently to  considerable mathematical complications. Our choice of a single overall $\FF$ is based on the following considerations.
Each phase is capable to move through a Lagrangian volume element but the mass of the mixture in the volume element remains constant. In other words, the Lagrangian coordinates are associated only with the overall mass (i.e. mass of the mixture)  but not with material particles of the components. In contrast, in the context of continuum mechanics of mixtures,  Lagrangian coordinates for individual components are commonly used (see for example Refs.~\cite{gav1999,cald,trusd}).
Our approach can be seen as a
natural generalization of the classical one-phase
Lagrangian dynamics that  also uses  only one set of
Lagrangian coordinates. The same approach  has already been  used in Refs.~\cite{rom2013,gav2011,favr2009}.

\subsection{\label{sec:model_lagr}Time evolution}

We continue with the  construction of  a particular realization of the Euler structure presented in Sections \ref{nvf} and \ref{dvf}. Our next task is to introduce equations governing the time evolution of the fields (\ref{stvnd}).  We shall write them  in the Cartesian
Lagrangian coordinates~$y_j$.  The requirement that such equations  possess the Godunov structure (\ref{xtcomp}) leads us to
\begin{subequations}\label{model_lagr}
\begin{align}
& \displaystyle\frac{{\rm d} v_i}{{\rm d} t} - \frac{\partial \mathscr U_{F_{i j}}}{\partial y_j} = 0,  \label{model_lagr_a}\\[2mm]
& \displaystyle\frac{{\rm d} F_{ij}}{{\rm d} t} - \frac{\partial \mathscr U_{v_{i}}}{\partial y_j} = 0,  \label{model_lagr_b}\\[2mm]
& \displaystyle\frac{{\rm d} c}{{\rm d} t} + \frac{\partial \mathscr U_{\hat{w}_j}}{\partial y_j} = -\hat{\chi},  \label{model_lagr_c}\\[2mm]
& \displaystyle\frac{{\rm d} \hat{w}_j}{{\rm d} t} + \frac{\partial \mathscr U_c}{\partial y_j} = -\hat{\eta}_j,  \label{model_lagr_d}\\[2mm]
& \displaystyle\frac{{\rm d} \alpha}{{\rm d} t}=-\hat{\theta},  \label{model_lagr_e}\\[2mm]
& \displaystyle\frac{{\rm d} S_l}{{\rm d} t}=\hat{\varsigma}_l, \ \ \ l=1,2, \label{model_lagr_f}\\[2mm]
& \displaystyle\frac{{\rm d} P_{ij}}{{\rm d} t}=-\hat{\Phi}_{ij}. \label{model_lagr_g}
\end{align}
\end{subequations}
The thermodynamic potential $\mathscr{U}(\qq)$ has the physical meaning of  the specific total energy of the mixture. Indeed, if the  partial entropies source terms $\hat{\varsigma}_l$ are taken in the form
\begin{equation}\label{entr_prod1}
\hat{\varsigma}_l=\frac{c_l}{\mathscr{U}_{S_l}}(\mathscr{U}_c\hat{\chi}+\mathscr{U}_{\hat{w}_j}\hat{\eta}_j+\mathscr{U}_\alpha\hat{\theta}+\mathscr{U}_{P_{ij}}\hat{\Phi}_{ij}), \ \ l=1,2,
\end{equation}
where $c_1=c$, $c_2=(1-c)$, then independently of the particular choice of $\mathscr{U}(\qq)$ (for its choice see
Section \ref{sec:plast}) and acording to the passage from (\ref{god_form1}) to (\ref{god_energy}) (or from (\ref{xtcomp}) to (\ref{Sod})) we see that 
\begin{equation}\label{lagr_en}
\frac{{\rm d}\mathscr{U}}{{\rm d}t}+\frac{\partial}{\partial y_j}\left(\mathscr{U}_c\mathscr{U}_{\hat{w}_j}-\mathscr{U}_{v_i}\mathscr{U}_{F_{ij}}\right)=0,
\end{equation}
which means that  the total energy $\int \mathscr{U} d\yy $ is conserved (the first law of thermodynamics).
In addition, in order to guarantee  the mathematical stability of (\ref{model_lagr}) or, in other words, in order (\ref{model_lagr}) be a symmetric hyperbolic system (\ref{eq:god_sym}), the potential $\mathscr{U}$ is required to be a convex function of $\qq$. 

We now demonstrate that Eqs. (\ref{model_lagr}) possess indeed  the Godunov structure provided the source terms appearing on the right hand side of (\ref{model_lagr}) satisfy (\ref{entr_prod1}). 


By a  direct verification,  we convince ourselves that (\ref{model_lagr}) can be indeed cast into the form (\ref{god_form1}) with
\begin{equation}\label{LML}
\begin{array}{l}
\pp=\mathscr{U}_{\qq},\\[3mm]
L(\pp)=v_i\mathscr{U}_{v_i}+F_{ij}\mathscr{U}_{F_{ij}}+c\mathscr{U}_{c}+\hat{w}_j
\mathscr{U}_{\hat{w}_j}+\alpha\mathscr{U}_\alpha+S_l\mathscr{U}_{S_l}+P_{ij}\scU_{P_{ij}}-\mathscr{U}=\qq^\mathsf{T}\pp-\mathscr{U}(\qq),  \\[3mm]
M^j(\pp)=\mathscr{U}_c\mathscr{U}_{\hat{w}_j}-\mathscr{U}_{v_i}\mathscr{U}_{F_{ij}}.
\end{array}
\end{equation}
and $\hat{\varsigma}_l$ given by (\ref{entr_prod1}).
This then implies in particular the extra conservation law (\ref{god_energy}) that in terms of $\qq$ and $\mathscr{U}$ reads as (\ref{lagr_en}).

In the matrix form, (\ref{model_lagr}) takes the form
\begin{equation*}
\frac{{\rm d} \qq}{{\rm d} t} + \frac{\partial \boldsymbol{\mathcal{F}}^j(\qq)}{\partial y_j} = \boldsymbol{\mathcal{S}}(\qq).
\end{equation*}
Here, ${\rm d}/{\rm d}t$ denotes the time-derivative along the  trajectory of a
fixed Lagrangian particle
\[\frac{{\rm d}}{{\rm d}t}\bigg|_{\yy ={\rm const}}=\frac{\partial}{ \partial t} +v_k\frac{\partial}{\partial x_k},\]
$\qq$ denotes the vector of conservative variables (\ref{stvnd}).
The symbol $\boldsymbol{\mathcal{F}}^j$ denotes  the  fluxes:
\[\boldsymbol{\mathcal{F}}^j=(-\mathscr U_{F_{i j}},-\mathscr U_{v_j},\mathscr U_{\hat{w}_j},\mathscr U_{c},0,0,\ldots,0)^\mathsf{T},\]
where the functions $\mathscr U_{F_{i j}}$ as well as $\mathscr U_{v_{i}}$, $\mathscr U_{\hat{w}_j}$ and $\mathscr U_c$ denote the partial derivatives of the total energy $\mathscr U$, i.e. $\mathscr U_{F_{i j}}\equiv\partial \mathscr U/\partial F_{i j}$, etc.,  $\boldsymbol{\mathcal{S}}$ denotes the dissipative vector field (i.e. the source terms).
The scalar functions $\hat{\chi}$ and $\hat{\theta}$ describe the phase
transition mechanism and the  process of equalizing the interfacial
pressure, respectively; the vector function $\hat{\boldsymbol{\eta}}=(\hat{\eta}_1,\hat{\eta}_2,\hat{\eta}_3)^\mathsf{T}$
describes an interfacial friction mechanism; the tensorial function (or dissipation tensor) $\hat{\boldsymbol{\Phi}}=[\hat{\Phi}_{ij}]$ describes a strain dissipation mechanism. By the symbol
$\hat{\varsigma}_l$ we denote  the entropy production in the phase $l$ due to
the dissipative processes. All the functions on the right hand side of
(\ref{model_lagr}) depend only on the unknown functions $\vv$, $\FF$,
$c$, $\hat{\ww}$, $\alpha$, $\PP$ and $S_l$ but not on their
spatial derivatives.

\paragraph{ \textbf{Remark 1.}} {\it We emphasize that the absence of dissipation in Eq. (\ref{model_lagr_b}) means that the dissipation does not directly influence macroscopic displacements (see (\ref{Lag})). This macroscopic motion is influenced by the dissipation only indirectly by influencing directly the internal structure characterized by $c$, $\alpha$, $\ww$ and $\PP$.}

\paragraph{\textbf{Remark 2.}} {\it We also could add to (\ref{model_lagr}) the mass conservation equation in the form
\begin{equation}\label{detF}
\frac{{\rm d} (\det\FF)}{{\rm d} t}-\frac{\partial }{\partial y_j} \left( H_{jk}v_k\det\FF\right)=0,
\end{equation}
where $\HH=\FF^{-1}=[H_{ij}]$, but such equation is clearly a consequence of equations (\ref{model_lagr}) and is thus unnecessary.}

\paragraph{ \textbf{Remark 3.}} {\it Finally, it is important to note (see also \cite{rom2001,god1996}) that the Godunov class of conservation laws, and in particular equations (\ref{model_lagr}), has a complementary structure implied by the summation rule (\ref{god_energy}). Namely, in order to satisfy (\ref{god_energy}) (or (\ref{lagr_en})),  the fields whose time evolution is governed by conservation laws with non-zero fluxes are grouped into pairs in which the second field in the pair has the physical interpretation closely related to the rate of the first field. This, of course, is reminiscent of the natural grouping of  state variables in the Hamiltonian systems (like position coordinates  and  corresponding to them velocities or momenta form one such group). For instance, the equations in (\ref{model_lagr}) are split  into the pairs (\ref{model_lagr_a}), (\ref{model_lagr_b}) and (\ref{model_lagr_c}), (\ref{model_lagr_d}). There are no restrictions on the number of equations that do not have fluxes, like e.g. (\ref{model_lagr_e}), (\ref{model_lagr_f}), (\ref{model_lagr_g}). This complementary structure will later be used (see Section \ref{sec:extension}) to derive an extension of the model in which more details of microscopic order will be involved.}

\subsection{Stress-based versus strain-based formulation}\label{sec: strain_stress}

In the classical rheological as well as plasticity models it is the stress tensor that plays the role of state variables. Such modeling is called stress-based. Our formulation presented above is strain-based since the role of the state variable is played by the strain tensor and the stress tensor arises as a quantity depending on the strain tensor and the remaining state variables.  The former modeling appears to be very straightforward  and natural if we think about comparison with experimental observations and practical application. This is because the state variable is a quantity that is directly measured. Indeed, at least in the classical continuum-physics measurements, it is the stress tensor (and not the strain tensor) that is directly measured. The latter (strain-base) modeling is however  more advantageous. Its advantages emerge in both physical and mathematical considerations.

As for the mathematical arguments, we have already seen them in the Lagrangian framework (\ref{model_lagr}) where the strain $\FF$ is used as the state variable and we shall also see them throughout this paper in other alternative formulations including Eulerian formulation and the formulations used in numerical calculations. The Godunov structure in which  the physical (i.e. in particular the compatibility with thermodynamics) and the mathematical regularity of the governing equations manifestly emerges is not seen in the stress based formulations.
Additional  arguments supporting the strain-based modeling will  also  arise in Section \ref{qvsp} and they also arise   in
the context of the Clebsch formulations  (see Refs.~\cite{G7,G8}).

The main physical argument  supporting the strain-based formulations is  the realization that it is the microscopic motion inside the system under consideration that determines its macroscopic behavior. The state variables (\ref{stvnd}) represent the mesoscopic representative of the complete microscopic characterization that would consist of  position and velocity coordinates  of all microscopic particles composing the system. The stress tensor is a quantity representing interactions with exterior of the system, it is not a quantity characterizing states of the system. It may happen that in some particular case the relation between stress tensor and strain tensor is one-to-one. In such particular case the stress-based and strain-based formulations are equivalent (but even in this case the strain-based formulation is preferable since the  governing equations are much simpler from the mathematical point of view). In general, the passage: strain $\rightarrow$ stress is a projection and consequently the strain-based formulation is the only choice. A very direct, experimentally based,  argument in favour of strain-based formulations is that the stress-based formulations are unable to predict the experimentally observed residual stresses mentioned in Section \ref{sec: Maxwell}.

\subsection{  \textit{\textbf{q}}-type variables  versus \textit{\textbf{p}}-type variables  }\label{qvsp}

The most important  contribution of macroscopic (or mesoscopic) physics (both static and dynamic) is the introduction of entropy. In the context of externally unforced systems, this new potential is required  to either remain constant (in nondissipative time evolution) or  reach its maximum allowed by constraints.  From the mathematical point of view, such maximization  is essentially a Legendre transformation. This then  implies immediately two types of state variables. One that are involved in the maximization (we shall call them $\qq$-type) and the other  that arise as Lagrange multipliers involved in the presence of constraints (we shall call them $\pp$-type, see (\ref{eq:p_q}) and (\ref{LML})). For example in classical thermodynamics, we have volume (a $\qq$-type variable) and corresponding to it pressure (a $\pp$-type variable), or similarly, energy and corresponding to it temperature. The most natural framework  for  Legendre transformations is the setting of contact geometry (in which the Clebsch based structure of mesoscopic dynamics is formulated in Ref.~\cite{G7,G8} ) involving  a large space in which  both $\qq$-type and $\pp$-type variables together with the potential that is maximized serve as independent state variables.  Their dependence, expressing the fundamental thermodynamic relation and thus, from the physical point of view, the individual features of the system under consideration, then takes the form of specification of a Legendre submanifold in the large space.

As we have already seen  in the previous section (see (\ref{eq:p_q}) or (\ref{LML})), the  $\qq$-type and the $\pp$-type variables as well as the Legendre transformations arise  naturally also  in the Godunov formulation of the Euler structure (see Sections~\ref{nvf} and \ref{dvf}). In this context, they are distinguished in addition by the fact that the time evolution of the $\qq$-type variables is governed by equations having the form of local conservation laws in which the  $\pp$-type variables appear in the fluxes. The time evolution of the $\pp$-type state variables themselves is not, in general,  governed by local conservation laws. In the finite volume discretization, needed in numerical calculations, this  distinction then implies that the $\qq$-type variables ``live'' inside the finite  volumes and the $\pp$-type variables on their boundaries.

The above comments about the duality of state variables provide also  additional arguments (in addition to those presented in Section \ref{sec: strain_stress}) in favour of  strain-based formulations.

\subsection{\label{sec:lagr_limit}Limitations of the Lagrangian framework}

If the mass fraction $1-c$  of the fluid phase is equal to zero or small enough, then the two phase mixture behaves essentially as an elastic or elasto-plastic solid and consequently   the Lagrangian framework appears to be  natural and  simple. If however the   deformations are large, the stress  is far above the yield limit,  and the fluid mass fraction is  large so that the two phase mixture exhibits predominantly fluid like motion,  then  the Lagrangian formulation  becomes inappropriate.  This is mainly because elements of both tensors $\FF$, $\PP$, that play the role of state variables in the Lagrangian formulation, experience, in general, an unlimited growth in the fluid like motion. Moreover, in numerical solutions of the governing equations the numerical mesh becomes strongly distorted and
complex remeshing procedures   (as for example, free-Lagrange numerical techniques or Arbitrary Lagrangian-Eulerean techniques) have to be applied.

Neither of these limitations applies however in the case of one dimensional systems to which we limit ourselves in numerical illustrations worked out in Section \ref{sec:num}. In general situations however, the limitations represent  a great obstacle and a special method is needed to overcome them.
We shall see below that
the Eulerian reformulation of  (\ref{model_lagr}) offers a convenient framework for describing  both solid like and fluid like behavior since only the elastic strain tensor $\EE$, that always remains bounded, appears as the state variable describing the strain.

\section{\label{sec:eul_dis}Euler Structure in the Eulerian Framework: Particular Realization for Solid-Fluid Mixture}

In this section we transform (\ref{model_lagr}) written in the Lagrangian (label) coordinates  $\yy$ into the Eulerian coordinates $\xx$ (these two types of coordinates are related by (\ref{Lag})).
Once the Eulerian equations will be  establish, we shall consider them in their own right. Their relation to the Lagrangian equations will be just one of their properties.

Before making the transformation, we recall an important  general fact about the relation between the Lagrangian and Eulerian frameworks. The Eulerian equations arise as a reduction of the Lagrangian equations by the group of symmetry consisting of relabeling the fluid particles. This viewpoint of the Lagrange-Euler relation  gets a very clear  mathematical formulation in particular in the framework of the  Clebsch structure formalism
(see e.g. Ref.~\cite{marsden,salmon}).

Since the passage Lagrange $\rightarrow$ Euler is a reduction, we cannot be surprised to loose in it a structure. For example,  the Hamiltonian structure of nondissipative Lagrangian hydrodynamic equations  is not lost in  the reduction but its canonical form appearing  in the Lagrangian framework transforms into a noncanonical and degenerate form in  the Eulerian framework. Below, we shall see that the Godunov structure of the Lagrangian equations does not survive in  its entirety  in the Lagrange $\rightarrow$ Euler passage.

\subsection{Eulerian equations}\label{eulnc}

We shall make the Lagrange $\rightarrow$ Euler transformation in this section directly by following Ref.~\cite{god1996} (some details are shown in Appendices~\ref{ap:momentum}, \ref{ap:rel} and \ref{ap:energy}). We use the notation introduced in (\ref{Lag}), i.e. $\yy$ are the Lagrangian coordinates and $\xx$ are the Eulerian coordinates. After straightforward calculations we arrive from
(\ref{model_lagr}) to
\begin{subequations}\label{model_eul}
\begin{align}
&\displaystyle\frac{\partial \rho v_i}{\partial t}+\frac{\partial (\rho v_i v_k+\rho^2\scE_\rho\delta_{ik}+\rho w_i \mathscr{E}_{w_k}+\rho A_{mi}\mathscr{E}_{A_{mk}})}{\partial x_k}=0,\label{model_eul_a}\\[2mm]
&\displaystyle\frac{\partial A_{i k}}{\partial t}+\frac{\partial A_{im} v_m}{\partial x_k}=-v_j\left(\frac{\partial A_{ik}}{\partial x_j}-\frac{\partial A_{ij}}{\partial x_k}\right)-\Phi_{ik},\label{model_eul_b}\\[2mm]
&\displaystyle\frac{\partial \rho c}{\partial t}+\frac{\partial (\rho c v_k+\rho \mathscr{E}_{w_k})}{\partial x_k}=-\chi,\label{model_eul_c}\\[2mm]
&\displaystyle\frac{\partial w_k}{\partial t}+\frac{\partial (v_m w_m + \mathscr{E}_c)}{\partial x_k}=-v_j\left(\frac{\partial w_{k}}{\partial x_j}-\frac{\partial w_{j}}{\partial x_k}\right)-\eta_k,\label{model_eul_d}\\[2mm]
&\displaystyle\frac{\partial \rho\alpha}{\partial t}+\frac{\partial \rho \alpha v_k}{\partial x_k}=-\theta,\label{model_eul_e}\\[2mm]
&\displaystyle\frac{\partial \rho S_l}{\partial t}+\frac{\partial \rho S_l v_k}{\partial x_k}=\varsigma_l,\ \ l=1,2,\label{model_eul_f}\\[2mm]
&\displaystyle\frac{\partial \rho }{\partial t}+\frac{\partial \rho v_k}{\partial x_k}=0.\label{model_eul_g}
\end{align}
\end{subequations}
Here, $\AAA=[A_{ij}]=\EE^{-1}$ is the inverse of the elastic strain, $\mathscr{E}$ is the total specific energy of the mixture. It relates to the energy $\mathscr{U}$ appearing in the Lagrangian framework as follow
\begin{equation}\label{eq:energy_E_U}
\scE(\vv,\AAA,c,\ww,\alpha,S_1,S_2,\rho)\equiv\scU(\vv,\FF,c,\hat{\ww},\alpha,S_1,S_2,\boldsymbol{P}).
\end{equation}
In particular, it is implied in (\ref{eq:energy_E_U}) that $\scE$ depends on $\FF$ and $\PP$ only through their dependence on $\AAA=\EE ^{-1}=\PP\FF^{-1}$.

The overall mass density $\rho$ has been defined in the Lagrangian framework by $\det\FF=\rho_0/\rho$, where $\rho_0$ is the reference mass density. Due to the way we define the strain dissipation mechanism in Section \ref{sec: Maxwell} (see condition~(D2)), this then also means that $\rho=\rho_0\det\AAA$. If we now multiply Eqs.~(\ref{model_eul_b}) by $\rho_{A_{ik}}$ (note that $\rho_\AAA=[\rho_{A_{ik}}]=\rho\AAA^{-\mathsf{T}}=\rho\EE^{\mathsf{T}}$) and sum them up we arrive at the mass conservation equation (\ref{model_eul_g}) (see Appendix~\ref{ap:energy})
This fact means that the mass conservation is a consequence of (\ref{model_eul_b}) and plays a role of the differential constraint for system (\ref{model_eul}). However, it simplifies the situation, e.g., numerical implementation of the model, if we include the density $\rho$ in the set of state variables for the reasons discussed in~\cite{miller}.

Summing up,
the state variables  $\qq$  that are used in  (\ref{model_eul}) are (compare with~(\ref{stvnd})):
\begin{equation}\label{eul_q}
\qq=(\rho\vv,\AAA,\rho c,\ww,\rho\alpha,\rho S_1,\rho S_2,\rho).
\end{equation}

The symbols with a hat, introduced in the Lagrangian framework, are related to the symbols without a hat by: $\hat{\chi}=\chi/\rho$, $\hat{\theta}=\theta/\rho$, $\hat{\boldsymbol{\eta}}=\FF^\mathsf{T}\boldsymbol{\eta}$, $\hat{\boldsymbol\Phi}=\boldsymbol\Phi \FF$,  and $\hat{\ww}=\FF^T\ww$. Finally, according to (\ref{entr_prod1}), (\ref{eq:energy_E_U}), and to the equalities $\scU_{\hat{\ww}}=\FF^{-1}\scE_\ww$, $\scU_\PP=\scE_\AAA\FF^{-\mathsf{T}}$, we define the entropy source terms as
\begin{equation}\label{eq:entr_prod_eul}
\varsigma_l=\frac{c_l}{\rho\,\scE_{S_l}}(\scE_c\chi+\rho\scE_{w_j}\eta_j+\scE_\alpha\theta+\rho\scE_{A_{ij}}\Phi_{ij}), \ \ l=1,2.
\end{equation}



\subsection{General properties of (\ref{model_eul})}

\subsubsection{Structure}

Our objective  is to cast (\ref{model_eul}) into the form (\ref{xtcomp}) of non-homogeneous Godunov conservation laws. We have succeeded to do it for (\ref{model_lagr}) but  in  (\ref{model_eul}) we shall be able to recognize only some elements of the Godunov structure. For example,
we  see clearly in (\ref{model_eul}) that this system of  equations does not even have  the form (\ref{xt}) of local conservation laws. Indeed, our first observation is that
the  transformation $\yy\rightarrow\xx$, in general, does not preserve the Godunov structure.
This observation was made first in the context of ideal magnetohydrodynamics equations \cite{god1972} (see also Ref.~\cite{god_rom2003}, p. 205). Among Eulerian models, it appears that the  hydrodynamics equations \cite{god1961} belong to the Godunov class of conservation laws only exceptionally. In the rest of this section we shall recognize in  (\ref{model_eul}) some parts of the Godunov structure (in particular we shall prove the thermodynamic compatibility) and investigate reformulations and extensions of (\ref{model_eul}) in which more elements of the Godunov  structure emerge.

\subsubsection{First law of thermodynamics}\label{sec:eul_therm_comp}

We begin with the energy conservation.  There are two routes that we can take.  We have already proven this result in the Lagrangian framework (see  (\ref{lagr_en})) and we can thus simply transform it into the Eulerian framework. The second route is to establish the compatibility with thermodynamics directly for (\ref{model_eul}) without any reference to the Lagrangian framework. We shall take the latter route.

Let the total energy density $\rho\mathscr E$ be a sufficiently regular function of $\qq$.
By  summing up of all equations (\ref{model_eul}) multiplied by the
corresponding multiplicative factors:
\begin{equation}\label{eul_p}
\pp=(\rho\mathscr E)_\qq=((\rho\mathscr E)_{\rho v_i},(\rho\mathscr E)_{A_{ik}},(\rho\mathscr E)_{\rho c},(\rho\mathscr E)_{w_i},
(\rho\mathscr E)_{\rho\alpha},(\rho\mathscr E)_{\rho S_l}, \scE - v_i\scE_{v_i}-c\scE_c-\alpha\scE_\alpha-S_l\scE_{S_l}-V\scE_V)^\mathsf{T},
\end{equation}
where $V=1/\rho$ is the specific volume,
we arrive at the  energy conservation equation (see Appendix~\ref{ap:energy})
\begin{equation}\label{eul_en}
\displaystyle\frac{\partial \rho \mathscr{ E}}{\partial t}+\frac{\partial }{\partial x_k}\left(v_k\rho \mathscr{E}+\rho \mathscr{E}_c
\mathscr{E}_{w_k}+\rho v_n (\rho\scE_\rho + w_n \mathscr{E}_{w_k}+ A_{mn}\mathscr{E}_{A_{mk}})\right)=0,
\end{equation}
if only the source terms $\varsigma_l$ for the partial entropies are given by~(\ref{eq:entr_prod_eul}).

The entropy conservation in the nondissipative time evolution is clearly visible in Eqs.~(\ref{model_eul_g}). Note, as in the Lagrangian framework (see (\ref{LML})), the variables $\qq$ and $\pp$ are conjugate: $\qq= L_\pp$,  $\pp=(\rho\mathscr E)_\qq$, where $L(\pp) = \qq^\mathsf{T}\pp-\rho\mathscr E$ is the Legendre transformation of the potential $\rho\mathscr{E}$.

\subsubsection{Mixture entropy and the second law}

In order to demonstrate the compatibility of (\ref{model_eul}) with the second law of thermodynamics,  we have to prove that for the dissipative part of the mixture entropy $s=c s_1 + (1-c)s_2$ time evolution the following inequality is satisfied
\begin{equation}\label{entr_prod2}
\left(\frac{\partial \rho s}{\partial t}\right)_{diss}=\left(\frac{\partial \rho S_1}{\partial t}\right)_{diss}+\left(\frac{\partial \rho S_2}{\partial t}\right)_{diss}=\varsigma_1+\varsigma_2\geq 0.
\end{equation}

We have already mentioned that particular choice (\ref{eq:entr_prod_eul})
\[\varsigma_l=\frac{c_l}{\rho\scE_{S_l}}(\scE_c\chi+\scE_{w_j}\eta_j+\scE_\alpha\theta+\scE_{A_{ij}}\Phi_{ij}), \ \ l=1,2.\]
for the partial entropies source terms gives the first law of thermodynamics~(\ref{eul_en}).  Hence, for the complete consistency with thermodynamics, it remains to show that our choice of the terms $\varsigma_l$ also satisfies the second law~(\ref{entr_prod2}).

Here, we restrict ourself only by mentioning the evident fact, i.e. that (\ref{entr_prod2}) is satisfied if 
the source terms $\chi$, $\eta_j$, $\theta$ and $\Phi_{ij}$ are defined in such a way  that they are  proportional to the multipliers $\scE_c$, $\scE_{w_j}$, $\scE_{\alpha}$, and  $\scE_{A_{ij}}$, respectively, with positive coefficients, i.e.
\begin{equation}\label{eq:source_term}
\chi \propto \scE_c, \ \ \ \eta_j\propto \scE_{w_j}, \ \ \  \theta\propto\scE_{\alpha}, \ \ \ \Phi_{ij}\propto\scE_{A_{ij}}.
\end{equation}
Of course, a particular specification of the total energy potential $\scE$ then should not be too restricted by this choice of the source terms but should provide sufficiently reach physics underlies system~(\ref{model_eul}). An example of the potential $\scE$ that seems capture sufficiently large physical details of dynamics of a solid-fluid mixture experiencing irreversible deformation (e.g. yield stress fluids) will be given in the following section, see (\ref{Ulagr}), (\ref{int_ener}), (\ref{eos_solid}), and (\ref{eos_fluid}).

\subsection{Specific properties of (\ref{model_eul})}

\subsubsection{\label{remark3}Energy}

Specific properties of solutions of (\ref{model_eul}) depend on the specific choice of the parameters (that are,  in general,   functions of the state variables) entering~(\ref{model_eul}). Most important among them is the energy $\scE$ (or $\scU$), since, as it is seen in (\ref{model_eul}) (or in (\ref{model_lagr})), it assumes the role of a generating potential (i.e. the fluxes  and also the source terms involve  derivatives of $\scE_\qq$, see the previous paragraph). How shall we express our physical insight into a solid-fluid mixture suffering irreversible deformation as a result of the solid-fluid phase transformation in~$\scE$?

We shall assume\cite{rom_drik2010} that  $\scE(\qq)$ is a sum of a specific internal energy $U(\AAA,c,\alpha,S_1,S_2,\rho)$ of the solid-fluid mixture, the specific kinetic energy $K(\vv)=\frac{\vv^{\mathsf{T}}\vv}{2}$ and the specific kinetic energy of relative motion $W(c,\ww)=c(1-c) \ww^{\mathsf{T}}\ww/2$:
\begin{equation}\label{Ulagr}
\scE(\qq)=U + K + W
\end{equation}
and \begin{equation}\label{int_ener}
U(\AAA,c,\alpha,S_1,S_2,\rho)=c\,U_1(\AAA,\rho_1,s_1)+(1-c)\,U_2(\AAA,\rho_2,s_2),
\end{equation}
where $U_1$ is the specific internal energy of  the pure solid phase ($c=1$) and $U_2$ is the specific internal energy of the pure  fluid phase ($c=0$). Moreover, we assume that  both $U_1$ and $U_2$ reach minimum when $\AAA=\boldsymbol{O}$, where $\boldsymbol{O}$ is an orthogonal tensor, or equivalently that
\begin{equation}\label{min}
U_\AAA=\boldsymbol{0}\ \ \ \mbox{if}\ \ \AAA=\boldsymbol{O}.
\end{equation}
A  specific choice of $U_1$ and $U_2$ will be made in numerical illustrations presented in Section~\ref{sec:phase_trans}.



\subsubsection{\label{sec:strlagr}Stress tensor}

As seen in Eq.~(\ref{model_lagr_a}),  the  first Piola-Kirchhoff stress tensor $\boldsymbol{\Pi}$ (non-symmetric, natural Lagrangian stress tensor) is given by
\begin{equation}\label{piola}
\boldsymbol{\Pi}=\mathscr{U}_{\FF}.
\end{equation}
This type of stress-strain relations is also referred to as hyperelastic type constitutive relations. According to (\ref{eq:energy_E_U}) and (\ref{Ulagr}), the Piola-Kirchhoff tensor $\mathscr{U}_{\FF}$ can be decomposed into the stresses appeared as the response to deformations and the stresses arisen as the response to relative motion (diffusion) of the two phases:
\begin{equation}\label{piola_decompose}
\mathscr{U}_{\FF} = U_\FF + W_\FF,
\end{equation}
where
\[W_\FF=[W_{F_{ij}}] = -(W_\ww^\mathsf{T}\otimes\ww)\HH^\mathsf{T} = -[H_{jm}w_i W_{w_m}].\]
Here, the notation $\HH=[H_{ij}]=\FF^{-1}$ and the formulae $W_{F_{il}}=W_{w_k}\partial
w_k/\partial F_{il}$ and $\partial H_{jk}/\partial
F_{il}=-H_{ji}H_{lk}$ or, in matrix notations, $\partial \HH/\partial \FF=-\HH^{\mathsf{T}}\otimes\HH$ are used. Symbol $\otimes$ denotes the Kronecker product.

In turn, after using the assumption (\ref{eq:energy_E_U}), $U_\FF$ can be written as
\begin{equation}\label{piola_int}
U_\FF = U_\EE\PP^{-\mathsf{T}},\ \ \EE=\AAA^{-1}.
\end{equation}
It is now clear that in order to compute the Piola-Kirchhoff stresses (\ref{piola}), we need to know two tensors among the three appearing in (\ref{FEP}). This is  the reason why it is necessary to consider the inelastic tensor $\PP$ as an independent state variable and  consequently include its time evolution  into (\ref{model_lagr}). In what follows, we shall see that in the Eulerian framework it  suffices to know only the tensor $\AAA$, or its inverse $\EE$, in order to compute the Cauchy stress tensor (natural Eulerian stress tensor).  The  time evolution equation for the tensors $\FF$ and $\PP$ are therefore absent in (\ref{model_eul}) (see also discussion on the limitation of the Lagrangian model in Section \ref{sec:lagr_limit}).

The Cauchy stress tensor $\TT=[T_{ij}]=\TT^\mathsf{T}$ appearing in momentum equation (\ref{model_eul_a}) is
\begin{equation}\label{total_stress}
\TT=\rho \FF\mathscr{U}_\FF^\mathsf{T}.
\end{equation}
This expression is directly derived during the Lagrange$\rightarrow$Euler transformation (see details in Appendix \ref{ap:momentum}). After (\ref{eq:energy_E_U}), in component form, (\ref{total_stress}) reads as
\begin{equation}\label{total_stress_comp}
T_{ik}=-\rho^2\scE_{\rho}\delta_{ik}-\rho w_i\mathscr E_{w_k}-\rho A_{mi}\mathscr{E}_{A_{mk}}\equiv -\rho^2 U_{\rho}\delta_{ik}-\rho w_i W_{w_k}-\rho A_{mi}U_{A_{mk}},
\end{equation}
where $\rho^2\scE_{\rho}=\rho^2 U_{\rho}=p$ is the mixture pressure.
If we however see the time evolution equations (\ref{model_eul}) in their own right (i.e. we do not think of them as being derived from  (\ref{model_lagr}) but as being proposed on the basis of a physical consideration of yield-stress fluids) then (\ref{total_stress_comp}) arises as a consequence of the requirement of energy conservation (\ref{eul_en}). Indeed, divergence form (\ref{eul_en}) of the energy evolution equation can not be derived from (\ref{model_eul}) in the manner of (\ref{god_energy}) (i.e. the total energy is not conserved) if $\TT$ is not given by (\ref{total_stress_comp}) (e.g. see p. 209 in Ref.~\cite{god_rom2003}). 

In order to obtain (\ref{total_stress_comp}) from (\ref{total_stress}), we use (\ref{Ulagr}), (\ref{piola_decompose}), and
\begin{equation}\label{rem2}
\rho\FF U_\FF^{\mathsf{T}}=\rho \FF (U_\EE \PP^{-\mathsf{T}})^{\mathsf{T}}=\rho\FF \PP^{-1} U_\EE ^{\mathsf{T}}=\rho\EE  U_\EE ^{\mathsf{T}},
\end{equation}
and that (e.g. see p. 69 in Ref.~\cite{god_rom2003})
\begin{equation}\label{sigmaE}
\rho \EE  U_\EE ^{\mathsf{T}} = -\rho \AAA ^{\mathsf{T}}U_\AAA.
\end{equation}





\subsubsection{\label{remark2}Strains}

We have seen that two  tensors (any two  among the three tensors $\FF,\EE,\PP$)  are needed  to compute stresses in the Lagrangian framework (see (\ref{piola_int})). The situation is  different in the Eulerian framework. As it is seen from (\ref{total_stress_comp}), (\ref{rem2}), and (\ref{sigmaE}), the  elastic strain $\AAA$ suffices to characterize the total stress tensor $\TT$. Consequently, there is no need to use the two strain tensors  $\FF$ and $\PP$ as state variables in the Eulerian framework. The Eulerian  equations  (\ref{model_eul}) are thus free from the limitations of the Lagrangian description mentioned earlier in Section~\ref{sec:lagr_limit}.

\subsection{Comments about the physical interpretation of (\ref{model_eul})}\label{sec:plast}


We proceed to discuss some  physical aspects of (\ref{model_eul}) that are more specifically  related to materials like yield-stress fluids. Illustrations of  numerical solutions of (\ref{model_lagr}) are worked out  in Section  \ref{sec:num}.


\subsubsection{Unloading and Strain Dissipation}\label{sec: Maxwell}

Let us now consider  one particular experimental observation  made on materials like yield-stress fluids, which is a typical example for materials, which can be represented by the solid-fluid continuum. The media, initially at rest,  is  subjected to an external stress (loading). Subsequently, the loading is removed and the material is let to evolve freely (unloading). The final state reached after removing the externally imposed stress is observed to be a state that is, in general, different from the initial state before the loading was applied~(incomplete recovery).

Consider the pure dissipative part of the overall time evolution (\ref{model_eul_b}), i.e.,
\begin{equation}\label{eq:time_evol_diss}
\left(\dfrac{\partial \AAA}{\partial t}\right)_{diss}=-\PHI,
\end{equation}
From the physical point of view, the dissipation tensor $\PHI$ represents changes in the internal structure, namely, rearrangement of structural elements. In view of (\ref{FEP}), this dissipation then results in preventing the complete recovery. 



Following Refs.~\cite{god_rom2003,god_rom1972,god1978}, we formulate mathematically the physical meaningfulness of the dissipation tensor $\boldsymbol{\Phi}$ by three conditions:
\begin{itemize}

\item[(D1)]
Shear stresses should relax during the strain dissipation process,

\item[(D2)]
determinant of $\AAA$ remains unchanged during the dissipative time evolution, $\left ( \frac{{\rm d} }{{\rm d}t} \det \AAA \right )_{diss}=0$,

\item[(D3)]
the production of the mixture entropy $s=c s_1 + (1-c) s_2$ is positive.
\end{itemize}

The first requirement is the well known Maxwell shear stress relaxation condition (e.g., see Ref.~\cite{god_rom2003}). The second requirement expresses the mass conservation, it can also be expressed  by the equality $\det \FF=\det \EE$, or $\det \PP=1$. Finally, the third condition guarantees that the entropy does not decrease. 


Our problem now is to find  at least some examples of $\boldsymbol{\Phi}$ satisfying the requirements (D1), (D2), (D3). In the following section, we show that if the strain dissipation function $\PHI$ is chosen to be proportional to $\scE_\AAA$ with a non-negative proportionality coefficient (see (\ref{eq:source_term})), then the three conditions are satisfied, which  emphasizes again the important role of the thermodynamic potential~$\scE$.

Subsequently, we discuss the incomplete recovery and the phenomenon of residual stresses. By applying the loading, the strain tensor $\FF$  changes. If the loading is removed then, in the absence of strain dissipation ($\FF=\EE$, $\PP=\II$), the material tends to recover its original shape (following the time evolution governed by Eqs.~(\ref{model_lagr_a}), (\ref{model_lagr_b})) characterized by the initial value of~$\FF$. Actually, the material will oscillate near the original stress free state. If however the source term ${\boldsymbol{\Phi}}$ in Eq.~(\ref{model_eul_b}) is different from zero (i.e. the strain dissipation is switched on) then $\FF\neq\EE$ and from (\ref{min}) and (\ref{piola_int}) it follows that a new equilibrium state is different from the initial state.



Now, we turn to  residual stresses. If $\AAA$ dissipates according to Eq.~(\ref{model_eul_b}), it dissipates, in general, in a different way in different locations. As a consequence, the state reached in the unloading process is, in general, strongly inhomogeneous and as such involves residual stresses which, in turn, can be connected with the aging of soft materials when properties of a sample (e.g., the value of yield stress, time relaxation, elastic moduli) depend on its history.

\subsubsection{\label{sec:phase_trans}Yielding and stress relaxation. Particular realisation for the micture internal energy}

Another experimental observation made on materials like yield-stress fluids is the occurrence  of yielding in which a solid like constitution changes into liquid like and solidification in which the same process but in the opposite  direction takes place.
This behavior is expressed in (\ref{model_eul}) in the dissipative terms $\boldsymbol{\Phi}$ and $\chi$  in Eqs.~(\ref{model_eul_b}) and (\ref{model_eul_c}), respectively.

Let us see first that a dissipation in the mass fraction $c$ has to be always accompanied with a dissipation in $\AAA=\EE^{-1}$. If $c=1$, the yield-stress fluid under consideration is an elastic solid and thus $\FF=\EE$. Let now $c$ decrease which means that the liquid phase starts to emerge. We recall that the tensors $\FF$ and $\EE$  address the motion of the overall mixture. In the presence of the liquid phase, the deformations characterized by $\FF$ cannot be anymore elastic since the liquid in the mixture will  not hold the stresses, i.e. the structural elements, those which are in the liquid state, will rearrange. This then means that the elastic distortion $\EE$, that is initially (i.e. when $c=1$)  equal to the total deformation gradient $\FF$, becomes   different from $\FF$, and this difference becomes  to play an important role in the time evolution. In other words, as the solid phase melts, the overall deformations become necessarily irreversible with $\EE\neq\FF$. The relaxation of tangential stresses in the strain dissipation processes (see (\ref{piola_int}) for the relation between the strain and the stress) is guaranteed by condition (D1) in Section~\ref{sec: Maxwell}. More details can be found in Ref.~\cite{god_rom2003}.
 Consequently, dissipative changes in $c$ have to always be  accompanied with dissipative changes in $\AAA$.

Now we turn to the mathematical formulations of the dissipative terms $\chi$ and $\boldsymbol{\Phi}$. We have already addressed  this question in Section \ref{sec: Maxwell}. From the physical point of view, the melting and solidifying process involved in the solid~$\leftrightarrow$~liquid transformations can be seen as a chemical reaction. The expression for  $\chi$ will thus arise in the mass-action-law formulation of its kinematics. In this paper we insist on considering only the time evolution that is compatible with thermodynamics. We therefore need a formulation of the mass-action-law dynamics that  is manifestly compatible with thermodynamics \cite{grmela2012}.

From the consequence (\ref{god_energy}) or (\ref{Sod}) of the Godunov structure and the requirement that the  entropy does not decrease during the dissipative time evolution we can conclude (see \ref{eq:entr_prod_eul}) that the  dissipative time evolution (denoted here by upper dot) of the partial entropies  $S_l$, $l=1,2$ is governed by (in this consideration we assume that $\ww=0$)
\begin{subequations}\label{eq:phase_trans}
\begin{align}
& \displaystyle \dot{S_l}=-\frac{c_l}{\scE_{S_l}}(\scE_c\dot{c}+\scE_\alpha\dot{\alpha}+\scE_{A_{ij}}\dot{A}_{ij}),\ \ l=1,2.\label{eq:phase_trans_a}\\[1.5mm]
& \displaystyle \dot{c}=-\chi/\rho,\\[2mm]
&\displaystyle \dot{\alpha}=-\theta/\rho, \\[2mm]
& \displaystyle \dot{\AAA}=-\PHI,
\end{align}
\end{subequations}
where $c_1=c$ and $c_2=1-c$. Consequently, in order to guaranty  $\dot{s}>0$, where  $s=S_1+S_2$ is the total entropy, it is sufficient to specify  $\dot{c}$, $\dot{\alpha}$, and $\dot{A}_{ij}$ in such a way  that their signs coincide with the signs of the multipliers $\scE_c$,  $\scE_\alpha$, and $\scE_{A_{ij}}$, respectively. With such a specification, all terms on the right hand side of (\ref{eq:phase_trans_a}) are positive. In addition, we should guaranty that our choice is consistent with conditions (D1), (D2), and (D3) from the previous section.

\paragraph*{\textbf{Example.\ }} \ \ \ As a simple illustration, we  choose  $\dot{c}=\frac{1}{\tau^{(c)}} \scE_c$ and $\dot{\alpha}=\frac{1}{\tau^{(\alpha)}} \scE_\alpha$ with positive functions $\tau^{(c)}$, $\tau^{(\alpha)}$. If we write explicitly the derivatives of the total  energy (\ref{Ulagr}) of the mixture with respect to the state variables $c$ and $\alpha$ (restricting ourselves to $\ww=0$), we obtain
\begin{equation}\label{p_explicit}
\begin{array}{c}
\displaystyle \scE_c=U_c+W_c=\mu_1-\mu_2\ \ \ \scE_\alpha=\frac{p_2-p_1}{\rho},\\[4mm]
\mu_l=U_l+p_l/\rho_l-s_lT_l,\ \ \ \displaystyle p_l=\rho_l^2\frac{\partial U_l}{\partial \rho_l},\ \ l=1,2,
\end{array}
\end{equation}
where $\mu_l$, $p_l$  and $T_l=U_{S_l}=\partial U_l/\partial s_l$ denote the chemical potential, the pressure and the temperature of the phase $l$ respectively. We thus see that changes of  $c$ in the dissipative time evolution result in changes  of $\alpha$ (due to dependence of pressures $p_l$, $l=1,2$  on $c$) and consequently the pressure equilibrium condition $p_1=p_2$ is violated. In addition,  variations of $c$, $\alpha$, and $\AAA$ make the entropies $S_l$, $l=1,2$  to change (see (\ref{eq:phase_trans})).




The strain dissipation tensor $\PHI$, internal energy $U$ and  relaxation times $\tau^{(c)}$, $\tau^{(\alpha)}$  remain still unspecified. In what follows, we specify them, prove that conditions (D1), (D2), and (D3) are fulfilled, and demonstrate a typical behavior of solutions to (\ref{eq:phase_trans}).

First, we specify the solid state internal energy  $U_1$ and the fluid phase internal energy $U_2$. We work under assumption that the internal energy  $U_1$ of the solid phase can be decomposed into two potentials, one  describing the hydrostatic and thermal energy density, $U_1^{hydro}(\rho_1,s_1)$, and the other  the contribution due to shear deformations $U_1^{shear}(\rho_1,s_1,I_1,I_2)$:
\[U_1=U_1^{hydro}+U_1^{shear},\]
where $I_1$, $I_2$ denote the invariants $I_1={\rm tr}(\AAA^\mathsf{T}\AAA)$, $I_2={\rm tr}(\AAA^\mathsf{T}\AAA)^2$.
In this work, we shall use
\begin{equation}\label{eos_solid}
\begin{array}{c}
U_1^{hydro}(\rho_1,s_1)=\dfrac{d_0^2}{\gamma(\gamma-1)}\left(e^{s_1/c_{V}}\left(\dfrac{\rho_1}{\rho_{01}}\right)^{\gamma-1}+ (\gamma-1)\dfrac{\rho_{01}}{\rho_1}\right),\\[5mm] U_1^{shear}=d_1^2\left(I_2-\dfrac{I_1^2}{3}\right)+\delta_Y,
\end{array}
\end{equation}
where $\rho_{01}$ is the initial mass density of the solid phase,  $c_{V}$ is the specific heat capacity at constant volume, $d_0$ and $d_1$ are the positive constants with the physical dimension of speed (in general, they are functions of the density $\rho_1$ and entropy $s_1$), $\gamma$ is the adiabatic exponent. The constant $\delta_Y$ is used to calibrate the mixture internal energy in order to make $\mu_1-\mu_2>0$ ($c$ decreases, fluidization) when stresses exceed the yield limit and $\mu_1-\mu_2<0$ ($c$ increases, solidification) if stresses are bellow the yield limit and $c<1$.


For consistency with the theory of linear elasticity, the constants $d_0$, $d_1$ are chosen as follows: $d_0 = \sqrt{c_l^2-4c_t^2/3}$, $d_1=c_t$, where $c_l$ and $c_t$ are the longitudinal and the transversal sound velocities of the solid phase at the reference stress-free configuration, respectively.

For the internal energy of the liquid phase,  we assume that it has the same expression as the hydrodynamic part $U_1^{hydro}$ of the solid internal energy, i.e.
\begin{eqnarray}\label{eos_fluid}
U_2(\rho_2,s_2) = \frac{d_0^2}{\gamma(\gamma-1)}\left(e^{s_2/c_{V}}\left(\frac{\rho_2}{\rho_{02}}\right)^{\gamma-1}+ (\gamma-1)\frac{\rho_{02}}{\rho_2}\right),
\end{eqnarray}
where $\rho_{02}$ is the reference mass density of the liquid phase. 

For a particular choice of $U$ given by  (\ref{eos_solid}) and (\ref{eos_fluid}), the shear part $\TT^{shear}=-\rho\AAA^\mathsf{T}(U_1^{shear})_\AAA$ of the total stress tensor $\TT$ is
\begin{equation}\label{eq:Tvisc}
\TT^{shear}=-4\,\rho\, d_1^2\GG\left(\GG - \dfrac{I_1}{3}\II\right)=-4\,\rho\, d_1^2\GG\left({\rm dev}\,\GG\right),\ \ \ \GG=\AAA^\mathsf{T}\AAA.
\end{equation}

Now we are in position to show that if the strain dissipation tensor $\PHI$ is chosen to be proportional to $\scE_\AAA=U_\AAA=(U_1^{shear})_\AAA$ with a non-negative proportionality coefficient, then three conditions (D1), (D2), and (D3) are satisfied.  We construct
\begin{equation}\label{eq:PSI}
\PHI = \dfrac{3}{4\, d_1^2\tau\Delta}\scE_\AAA=\dfrac{3}{\tau\Delta}\AAA ({\rm dev}\,\GG),\ \ \ \Delta=\det\AAA>0,
\end{equation}
where $\tau$ is the characteristic time of strain dissipation.

It is clear that condition (D3) is automatically satisfied if $\PHI$ is given by formula (\ref{eq:PSI}) because the right-hand side of (\ref{eq:phase_trans}) becomes a quadratic form with positive coefficients. By comparing (\ref{eq:Tvisc}) and (\ref{eq:PSI}), it is obvious that condition (D1) is also satisfied. A proof of condition (D2) can be found in~\cite{god_rom2003}.

Second, we specify relaxation times $\tau^{(c)}$, $\tau^{(\alpha)}$, and $\tau$. For simplicity we assume that $\tau^{(\alpha)}=0$, i.e. the phase pressure equalizing velocity is infinite. In this case the third ordinary differential equation in (\ref{eq:phase_trans}) is substituted by the algebraic equation $p_1-p_2=0$.

Next, we assume that
\begin{equation}\label{chi}
\dfrac{1}{\tau^{(c)}}=\frac{1}{\tau_1}+\frac{1}{\tau_2},\ \ \ \displaystyle\tau_1=\tau_{01}\exp(n_1(\sigma_Y-\sigma_I)),\ \ \ \tau_2=\tau_{02}\exp(n_2(\sigma_I-\sigma_Y)).
\end{equation}
The symbol $\tau_1$ denotes  the characteristic time of the solid-to-fluid transition (it can also be seen as the characteristic time of the bond destruction), $\tau_2$ is the characteristic time of the fluid-to-solid transition  (the characteristic time of the bond formation). The quantities $\tau_{01},\tau_{02},n_1,n_2$ are positive material parameters, $\sigma_Y$ is the yield stress, and $\sigma_{I}=\left((\sigma_1-\sigma_2)^2+(\sigma_2-\sigma_3)^2+(\sigma_3-\sigma_1)^2\right)^{1/2}$ is the intensity of tangential stresses; $\sigma_1,\sigma_2$, and $\sigma_3$ are the eigenvalues of the stress tensor  (\ref{sigmaE}). Note also  that if $\AAA$ has the singular values $a_i$ then the principal stresses $\sigma_i$  are given by $\sigma_i=\rho_0a_1a_2a_3(a_iU_{a_i})$. Thus, the characteristic times $\tau_1$ and $\tau_2$ are functions of the singular values~$a_i$. In addition, the equation~(\ref{eq:PSI}) provides that the distortion $\AAA$ and dissipation function $\PHI$ are coaxial. This means that the nine ordinary differential equations $\dot{\AAA}=-\PHI$ equivalent to the following three differential equations:
\begin{equation}\label{eq:time_evol_sing}
\frac{\partial a_i}{\partial t}=-\frac{(2\,a_i^2 - a_m^2 - a_n^2)}{\tau a_m a_n},\ \ \ i\neq m\neq n\neq i,
\end{equation}
written in the terms of singular values $a_i$.

Finally, we specify the strain dissipation time $\tau$ in (\ref{eq:PSI}) as
\begin{equation}\label{tau}
\tau=\left\{
\begin{array}{ll}
\infty, & {\rm if}\ c=1,\\[4mm]
\displaystyle\tau_0\exp\left(-\frac{1}{c^m}\right), &{\rm if}\ c<1,
\end{array}\right.
\end{equation} where $\tau_0$, $m$ are positive material parameters. Equation ~(\ref{tau}) defines a monotonic function $\tau(c)$ that  tends to 0 as $c\rightarrow 0$  and  equals  $\infty$ if  $c= 1$. In other words, the velocity of the strain relaxation, $1/\tau(c)$, tends to infinity in the liquid limit and equals to zero in the solid limit.  Note that actually $c$ cannot take the values 1 or 0, because the model degenerates at this values, and in computations bellow $c$ takes values in the interval $0<\varepsilon\leq c \leq 1-\varepsilon$ for some small $\varepsilon$. The same valid for the volume fraction $\alpha$.

The complete  system of ordinary differential equations governing the dissipative time evolution (with  $\ww=0$) is given by
\begin{subequations}\label{eq:phase_trans1}
\begin{align}
& \displaystyle \dot{c}=-\frac{1}{\tau^{(c)}}\scE_c,\\[2mm]
&\displaystyle \dot{\alpha}=-\frac{1}{\tau^{(\alpha)}}\scE_\alpha, \\[2mm]
& \displaystyle \dot{\AAA}=-\dfrac{3}{4\, d_1^2\tau\Delta}\scE_\AAA,\\[2mm]
& \displaystyle \dot{S_l}=-\frac{c_l}{\scE_{S_l}}(\scE_c\dot{c}+\scE_\alpha\dot{\alpha}+\scE_{A_{ij}}\dot{A}_{ij})\geq 0,\ \ l=1,2.
\end{align}
\end{subequations}

\begin{figure}
\includegraphics[trim = 0mm 60mm 0mm 70mm, clip, scale=0.5]{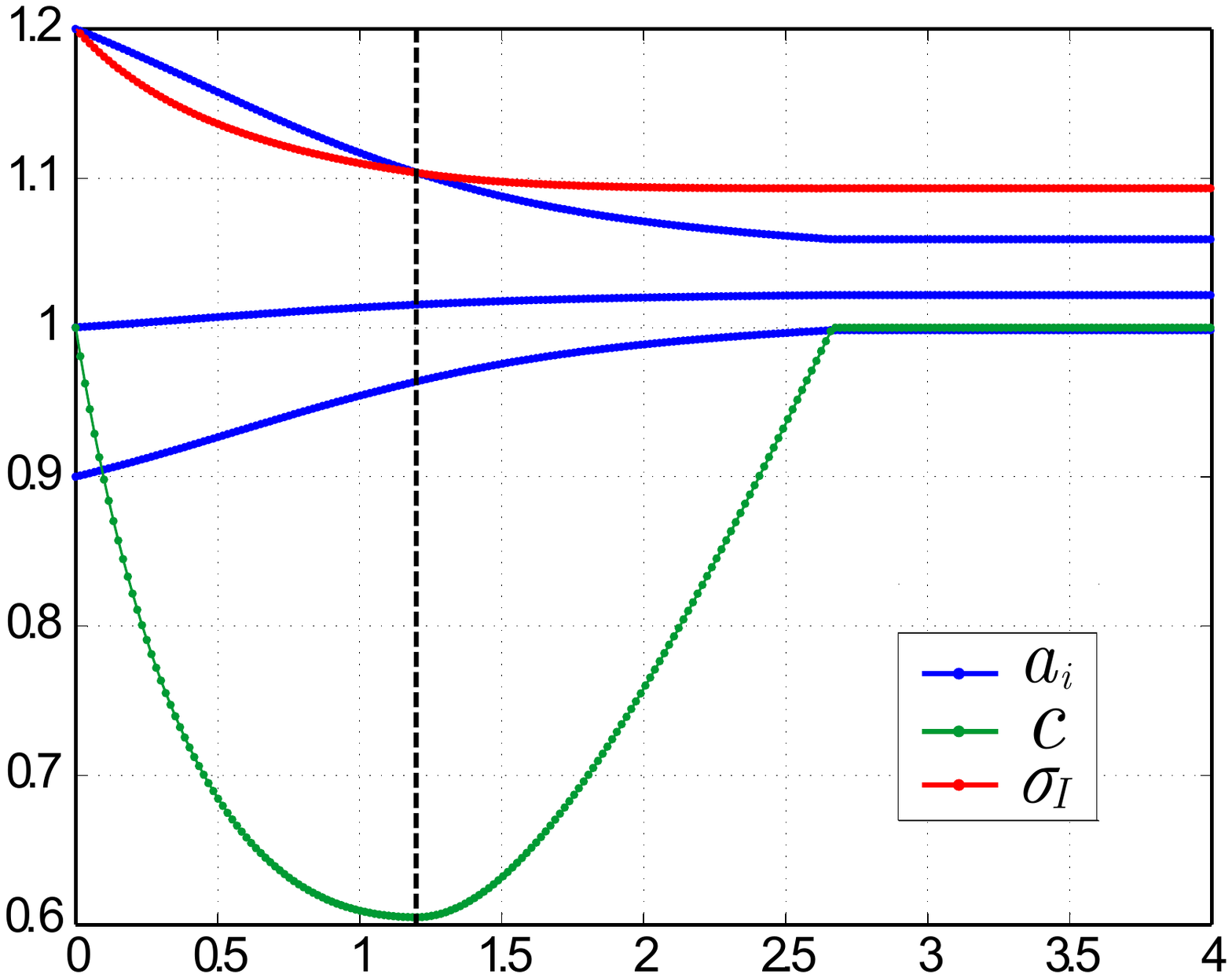}
\caption{\it Qualitative behavior of the singular values $a_i$, the mass fraction $c$ and the intensity of tangential stresses $\sigma_I$ in the solid-fluid transition. }\label{fig:a_c}
\end{figure}

A typical behavior of the mass fraction $c$ and singular values $a_i$ governed by dissipative time evolution (\ref{eq:phase_trans1}) is presented on Fig.~\ref{fig:a_c} (it appears that for $\tau^{(\alpha)}=0$  and for the given $U_1$ and $U_2$ we have the equality $\alpha=c$).

Fig.~\ref{fig:a_c} depicts numerical solution to (\ref{eq:phase_trans1}) with the initial data $a_1=1.2$, $a_2=1$, $a_3=0.9$, $c=0.99999$. The other parameters in (\ref{eq:phase_trans1}) are: $\delta_Y=-0.01$ that corresponds to the yield limit $\sigma_Y=0.0275$ GPa, $\tau_0=20$ sec$^{-5}$, $\tau_{01}=\tau_{02}=0.01$ sec$^{-5}$, $m=m_1=1$, $m_2=0.05$, $n_1=n_2=1$. In addition, the intensity of tangential stresses $\sigma_I$ corresponding to the initial data equals to 0.13 GPa. Since we use $\sigma_I > \sigma_Y$ in the initial condition,  the material starts to transform into a fluid  and eventually yields. The vertical dashed line on Fig.~\ref{fig:a_c} denotes the moment of the time when the intensity of the tangential stresses $\sigma_I$ (red line) becomes equal to $\sigma_Y$. At that moment fluidization process switch to the solidification one (green line). For the sake of convenience, we lift the curve $\sigma_I$ by adding~1.0677.

It is important to remark that the strain dissipation process has the same direction on either side of the vertical dashed line. This happens because the material in the fluid state appears  on the both sides which then means that the microstructure continues to rearrange. Consequently, the singular values $a_i$ are relaxing at the both sides of the dashed line while the mass fraction of the solid phase $c$ decreases (fluidization) on the left side of the vertical line and it increases (solidification) on the right. We note that despite such antagonistic actions of the dissipative processes depicted in  the right side of Fig.~\ref{fig:a_c} (i.e. their influences on elastic modulus are different;  the solidification process tends to increase shear modulus while the strain dissipation tends to relax them) the shear stress $\sigma_I$ (red line on Fig.~\ref{fig:a_c}) remains almost at the same level after it has reached the value of the yield stress at the moment of time denoting by the vertical solid line.

We conclude this illustration by a brief discussion of the  way how to fit  solutions of the presented model to experimental data. It is clear that the non-dissipative part (i.e. the left hand side of (\ref{model_lagr})) of the total time evolution (\ref{nd}) are rather universally valid.  The key parameters of the model that distinguish one material from another appear in the  dissipative source terms (i.e. on the right hand side of (\ref{model_lagr})). In particular, in the case of an irreversible deformation of a solid-fluid mixture,  the key parameters are the phase transition characteristic times $\tau_1$,  $\tau_2$ (see (\ref{chi})) and the strain dissipation characteristic time $\tau$ (see (\ref{tau})). In order to fit experimental data, one has to therefore specify first these parameters. Above, we suggested a basic form for these key functions. In general, one can simulate different non-Newtonian properties such as thixotropy, rheopecty, shear thickening and shear thinning by choosing proper expressions for the functions $\tau_1$, $\tau_2$ and $\tau$.



\subsection{\label{eulc}Reformulations of the Eulerian equations}

In this subsection we  begin to explore the structure of Eqs.~(\ref{model_eul}). We know that the Godunov structure of the time evolution equations  implies an important information about their solutions. Equations~(\ref{model_eul}) possess only some elements of this structure. We shall therefore begin to search for reformulations of Eqs.~(\ref{model_eul}) in which a richer structure, that is closer to  the complete Godunov structure, emerges. Even if this exploration is incomplete, it has an interesting physical content, it provides a new look on the governing equations (\ref{model_eul}), it may serve as a guide to subsequent explorations,  and,  in view of the importance of the Godunov structure in establishing the mathematical and numerical regularity of the governing equations.

We begin with  a reformulation that brings (\ref{model_eul}) into a system
of local conservation laws.

Our starting point is the following observation.  By applying  $\varepsilon_{jkl}\partial/\partial x_l$, where $\varepsilon_{jkl}$ is the unit pseudoscalar,  on  (\ref{model_eul_b}) and (\ref{model_eul_d}) we arrive  \cite{rom_tor2004,favr2009} at
\begin{subequations}\label{eul_constr}
\begin{align}
&\displaystyle\frac{\partial B_{ij}}{\partial t}+\frac{\partial }{\partial x_k}\left(v_k B_{ij}-v_j B_{ik}+\varepsilon_{jm k}\Phi_{im}\right)=0,\label{eul_constr2_a}\\[2mm]
&\displaystyle\frac{\partial \omega_j}{\partial t}+\frac{\partial }{\partial x_k}\left(v_k \omega_j-v_j \omega_k+\varepsilon_{jm k}\eta_m\right)=0,\label{eul_constr2_b}
\end{align}
\end{subequations}
where
\begin{equation}\label{eq:burg}
\BB =[B_{ij}]={\rm rot}\AAA \equiv{\rm rot}\EE ^{-1}
\end{equation}
i.e. $B_{ij}=\varepsilon_{jlk}\frac{\partial A_{il}}{\partial x_k}$  and
\begin{equation}\label{EO}
\boldsymbol\omega=(\omega_1,\omega_2,\omega_3)^\mathsf{T}={\rm rot} \ww
\end{equation}
i.e. $\omega_j=\varepsilon_{jlk}\frac{\partial w_l}{\partial x_k}.$
This means that the dissipative Eulerian model (\ref{model_eul}) is an overdetermined system of partial differential equations. Namely, the fields $\AAA$ and $\ww$ have to satisfy not only the time evolution equations ~(\ref{model_eul_b}) and (\ref{model_eul_d}) but also to stationary conservation laws (\ref{eq:burg}) and (\ref{EO}) with  $\BB$ and $\boldsymbol{\omega}$ obeying to non-stationary conservation laws  (\ref{eul_constr2_a}) and (\ref{eul_constr2_b}).

We note that the new quantities $\BB$ and $\boldsymbol\omega$ introduced in (\ref{eul_constr}) have a clear physical meaning.

First, we turn to the vector ${\boldsymbol\omega}$. If $\ww$ in (\ref{EO}) is replaced by the overall velocity $\vv$ then ${\boldsymbol\omega}$ is a well known vorticity vector. Consequently, ${\boldsymbol\omega}$ is a vorticity corresponding to the vector of the relative velocity $\ww$.

Next, we note that the  tensor $\BB$ is known in the plasticity theory  (see for example Refs.~\cite{god_rom2003,steinmann})  as the Burgers tensor or also as the dislocation density tensor.
We shall discuss its possible microscopic or mesoscopic interpretations for amorphous materials like yield-stress fluids later in this section.

In the rest of this section we shall investigate some consequences of (\ref{eul_constr}).

\subsubsection{Gauge constraint in the nondissipative time evolution}

If the  dissipation is absent, i.e. if $\boldsymbol{\Phi}=0$ and $\boldsymbol\eta=0$ in (\ref{eul_constr}),  then (\ref{eul_constr}) implies that the equalities
\begin{equation}\label{eul_constr2}
    {\rm rot}\AAA =0, \ \ \ {\rm rot}\ww =0
\end{equation}
hold for all times $t>0$  provided they  hold for the initial time $t=0$. The equalities (\ref{eul_constr2}) represent thus a constraint  that we shall refer to as a \textit{gauge constraint}.

We make a few observations.

\paragraph{ \label{observ1}\textbf{Observation 1.}}\ First, we note that the discovery of the gauge constraint (\ref{eul_constr2}) is an important  result about solutions of (\ref{model_eul}). Indeed, the system of hyperbolic type equations (\ref{model_eul}) with no source terms  is shown to be in fact an overdetermined system of equations coupled to the constraint (\ref{eul_constr2}) that is of the elliptic type. Overdetermined systems of hyperbolic-elliptic type are well known in computational continuum mechanics mainly due to the fact that   inevitable errors in numerical calculations cause the
constraints to be violated and consequently  numerical solutions become physically meaningless  unless  some sophisticated constraints-treatment procedures (see for example Refs.~\cite{powell,munz,miller,babii} and references therein) are implemented. This difficulty is one of the reasons why we shall attempt (in the next section) to lift (\ref{model_eul})  to a larger system that is free of constraints and  all equations in the system are local conservation laws.

\paragraph*{ \label{observ2}\textbf{Observation 2.}}\ An interesting question is  how does the gauge constraint appear and what is its role in the Lagrangian framework. We leave this question here  without an answer. We hope to investigate it in a future paper.

\paragraph*{ \label{observ3}\textbf{Observation 3.}}\ We recall that the gauge constraint  (\ref{eul_constr2}) holds when the dissipation is absent. But in such case also $\PP=\II$  (i.e. the deformations are reversible) and thus the Euler coordinates $\xx$ and Lagrange coordinates $\yy$ are related by
$$A_{jk}(x_1,x_2,x_3)=\frac{\partial y_j}{\partial x_k}.$$   As it is well known in  elasticity theory (see for example Ref.~\cite{god_rom2003}),  ${\rm rot}\AAA =0$ guarantees  that this system of equations has a unique solution $\yy(\xx)$.

\paragraph*{ \label{observ4}\textbf{Observation 4.}}\  So far, we have seen that (\ref{model_eul}) without the source terms (i.e. without dissipation) and with the gauge constraints (\ref{eul_constr2}) (stationary conservation laws) constitutes  a system of local conservation laws implying an additional conservation law (namely the conservation of energy) but, because of the presence of the gauge constraint, the system as such cannot be symmetrized and thus its mathematical regularity remains open.

Following to \cite{god1972,rom2001,god_rom2003}, we now present  a reformulation of (\ref{model_eul}) that violates the property that all  equations are local conservation laws but that is free of constraints (\ref{eul_constr2}) (i.e. all solutions of the reformulated model will automatically satisfy (\ref{eul_constr2})) and admits the  symmetrization.
We note that by adding
\begin{equation}\label{eul_0}
\rho\mathscr{E}_{A_{jk}}\left(\frac{\partial A_{ik}}{\partial x_j}-\frac{\partial A_{ij}}{\partial x_k}\right)+
\rho\mathscr{E}_{w_{j}}\left(\frac{\partial w_{k}}{\partial x_j}-\frac{\partial w_{j}}{\partial x_k}\right)
\end{equation}
to the momentum equation (\ref{model_eul_a}) (due to the constraints (\ref{eul_constr2}) both terms in (\ref{eul_0}) equal zero and we therefore do not change (\ref{model_eul_a})) the system of equations (\ref{model_eul}) can be cast into the form
\begin{equation}\label{eul_sym}
\begin{array}{ccccl}
\displaystyle\frac{\partial L_{v_i}}{\partial t} & + & \displaystyle\frac{\partial M^k_{v_i}}{\partial x_k} & + & \displaystyle L_{\alpha_{im}}\frac{\partial \alpha_{km}}{\partial x_k}
-L_{\alpha_{mk}}\frac{\partial \alpha_{mk}}{\partial x_i}+L_{\kappa_i}\frac{\partial \kappa_k}{\partial x_k}-L_{\kappa_{m}}\frac{\partial \kappa_{m}}{\partial x_i}=0,\\[5mm]
\displaystyle\displaystyle\frac{\partial  L_{\alpha_{il}}}{\partial t} & +  & \displaystyle \frac{\partial M^k_{\alpha_{il}}}{\partial x_k} & + & \displaystyle L_{\alpha_{ml}}\frac{\partial v_m}{\partial x_i}
-L_{\alpha_{il}}\frac{\partial v_k}{\partial x_k}=0,\\[5mm]
\displaystyle\frac{\partial L_\nu}{\partial t} & + & \displaystyle\frac{\partial M^k_\nu}{\partial x_k} & + & \displaystyle\frac{\partial \kappa_k}{\partial x_k}=0,\\[5mm]
\displaystyle\frac{\partial  L_{\kappa_i}}{\partial t} & + & \displaystyle\frac{\partial M^k_{\kappa_i}}{\partial x_k} & + & \displaystyle L_{\kappa_{m}}\frac{\partial v_{m}}{\partial x_i}
-L_{\kappa_i}\frac{\partial v_k}{\partial x_k}+\frac{\partial \nu}{\partial x_i}=0,\\[5mm]
\displaystyle\frac{\partial  L_{\gamma_i}}{\partial t} & + & \displaystyle\frac{\partial M^k_{\gamma_i}}{\partial x_k} & = & 0,\ \ i=1,2,3,4
\end{array}
\end{equation}
where $L(\pp)$ is the Legendre transformation of $\rho\mathscr{E}$ with respect to $\pp$ which is given in (\ref{eul_p}), we use here the notation: $\pp=(\vv,\boldsymbol{\alpha}, \nu, \boldsymbol{\kappa}, \gamma_1,\gamma_2,\gamma_3)$, $\boldsymbol{\alpha}=[\alpha_{ij}]=[(\rho\mathscr{E})_{A_{ij}}]$, $\nu=(\rho\mathscr{E})_{\rho c}$, $\boldsymbol{\kappa}=(\kappa_1,\kappa_2,\kappa_3)^\mathsf{T}=((\rho\mathscr{E})_{w_1},(\rho\mathscr{E})_{w_2},(\rho\mathscr{E})_{w_3})^\mathsf{T}$, $\gamma_1=(\rho\mathscr{E})_{\rho S_1}$, $\gamma_2=(\rho\mathscr{E})_{\rho S_2}$, $\gamma_3=(\rho\mathscr{E})_{\rho\alpha}$, $\gamma_4=\scE - v_i\scE_{v_i}-c\scE_c-\alpha\scE_\alpha-S_l\scE_{S_l}-V\scE_V$; moreover, $M^k(\pp)=v_kL(\pp)\,$.
If we keep in this system only the first two terms then we have the system in the Godunov form (\ref{god_form1}) that, as we have seen, can be symmetrized (see (\ref{eq:god_sym})). The remaining terms violate the conservative form but, as a direct verification shows, contribute only by adding to the symmetric matrix $M^k_{\pp\pp}$ appearing in (\ref{eq:god_sym}) another symmetric matrix. We have thus proven that the Cauchy problem for the nondissipative version of (\ref{model_eul}) is well posed.

A generalization of the concept of thermodynamically consistent  conservation laws with gauge constraints (e.g. (\ref{eul_constr2})) in the Eulerian framework is  the subject of  the series of papers \cite{god1995,god1996,rom1998,rom2001,god_rom2003}. We shall discuss these issues  below in this section.

\subsubsection{Extended system of dissipative time evolution equations}\label{sec:extension}


We  now consider the general case (i.e. the case when the dissipation is included and thus    $\boldsymbol\Phi\neq 0$ and/or  $\boldsymbol\eta\neq 0$)  and attempt to reformulate  (\ref{model_eul}) into a system of local conservation laws free of gauge constraints like (\ref{eul_constr2}). The main idea of the reformulation is an extension of the set of the state variables $\qq$ which is consistent with the constraints (\ref{eul_constr2}). Below,  we shall suggest such extension. Our objective is  to present   the main idea of the extension  and to draw attention on the way the conservation principle may help to derive new models of complex media. We hope to follow this line of research in a future publication.


For the sake of simplicity, we shall illustrate the extension only for the state variable $\AAA$. We shall also omit the equations for the variables $c$, $\alpha$, $S_l$ since they remain unchanged. This means that we begin with the set of state variables $\qq=(\rho\vv,\AAA)$ and the time evolution equations (\ref{model_eul_a}), (\ref{model_eul_b}). The extended set of state variables will be denoted $\qq^{(ext)}=(\qq,\widetilde{\qq})$. The first question that arises is of how do we choose $\widetilde{\qq}$.

We can investigate this question on the mathematical and the physical grounds. The former type of investigation leads us to the choice $\widetilde{\qq}=\BB$, where $\BB$ is Burgers tensor (\ref{eq:burg}). This indeed follows from the observations about the mathematical structure of (\ref{model_eul})  that we have made in the previous section. The physical arguments supporting this choice are based on the requirement that the extension that we are making is physically meaningful in the sense that it reaches to a more microscopic description in which more microscopic details (but only those of essential importance for the problem under investigation) are taken into account. The Burgers tensor $\BB$ characterizes indeed the microstructure, specifically,
it characterizes the defect distribution in the material (e.g.,
dislocation density in crystalline solids). The presence of defects implies that the character of
interactions among structural elements  differ from the one in a
defect free state. In other words, an infinitesimal
volume ${\rm d}\zz=\PP{\rm d}\yy$ of the material in the intermediate
configuration has, in general, different mechanical properties (e.g. yield stress, elastic
modulus, characteristic time of stress relaxation) than its inverse image ${\rm d}\yy=\PP ^{-1}{\rm d}\zz$
in the reference configuration has. The Burgers tensor is thus needed to take this fact into account \cite{steinmann,god2004,rom_sad}. In addition, the theory of flow defects \cite{langer} provide other arguments supporting the physical significance of the tensor $\BB$ (and also the tensor $\DD$ introduced below in (\ref{eul_final})).

If we decide to consider the Burgers tensor $\BB$ as an independent state variable we  see immediately  that we need another vector valued field, that we denote $\ff=(f_1,f_2,f_3)^\mathsf{T}$, to also admit  as an independent state variable.  This is because $\BB$, even if seen as an independent state variable, is itself constrained by  its origin, namely by the fact that $\BB$  is a rotation of a tensor. This then automatically  implies  constraint   ${\rm div} \BB=0$ since the operation of rotation followed by divergence leads always to zero. In order to take into account this new constraint we need a new vector $\ff={\rm div} \BB$, i.e. $f_i=\partial B_{ik}/\partial x_k$. Summing  up, we have introduced $\qq^{(ext)}=(\rho\vv,\AAA,\BB,\ff)$. We suggest now that the time evolution of $\qq^{(ext)}$ is governed by non-homogeneous local conservation laws (extending Eqs.~(\ref{model_eul_a}) and (\ref{model_eul_b}) ) of the following form:
\begin{subequations}\label{eul_num1}
\begin{align}
&\displaystyle\frac{\partial \rho v_i}{\partial t}+\frac{\partial (\rho v_i v_k + \rho^2\scE_\rho \delta_{ik} + \rho A_{mk}\mathscr{E}_{A_{mi}})}{\partial x_k}=0,\label{eul_num1_a}\\[2mm]
&\displaystyle\frac{\partial A_{i k}}{\partial t}+\frac{\partial A_{im} v_m}{\partial x_k}=-\varepsilon_{kml}v_m B_{il}-\Phi_{ik},\label{eul_num_b}\\[2mm]
&\displaystyle\frac{\partial B_{ij}}{\partial t}+\frac{\partial }{\partial x_k}\left(v_k B_{ij}-v_j B_{ik}+\varepsilon_{jm k}\Phi_{im}\right)=-v_jf_i,\label{eul_num_c}\\[2mm]
&\displaystyle\frac{\partial f_i}{\partial t}+\frac{\partial f_i v_k}{\partial x_k}=0.\label{eul_num_d}
\end{align}
\end{subequations}

This extension is still  however incomplete. The local conservation laws (\ref{eul_num1}) do not imply the energy conservation since (\ref{eul_num1}) do not possess the complementary structure described at the last paragraph of Section~\ref{sec:model_lagr}. We suggest therefore to continue the extension and consider  $\qq^{(ext)}=(\rho\vv,\AAA,\BB,\DD,\ff,\boldsymbol{g})$ as the set of fields representing independent state variables. The time evolution equations take now  the form
\begin{subequations}\label{eul_final}
\begin{align}
&\displaystyle\frac{\partial \rho v_i}{\partial t}+\frac{\partial (\rho v_i v_k + \rho^2\scE_\rho \delta_{ik} -\rho B_{mk}\mathscr{E}_{B_{mi}}-\rho D_{mk}\mathscr{E}_{D_{mi}}+\rho A_{mk}\mathscr{E}_{A_{mi}})}{\partial x_k}=0,\label{eul_final_a}\\[2mm]
&\displaystyle\frac{\partial A_{i k}}{\partial t}+\frac{\partial A_{im} v_m}{\partial x_k}=-\varepsilon_{kml}v_m B_{i l}-\Phi_{ik},\label{eul_final_b}\\[2mm]
&\displaystyle\frac{\partial B_{ij}}{\partial t}+\frac{\partial }{\partial x_k}\left(v_k B_{ij}-v_jB_{ik}+\varepsilon_{jm k}\mathscr{E}_{D_{im}}\right)=-v_j f_i,\label{eul_final_c}\\[2mm]
&\displaystyle\frac{\partial D_{ij}}{\partial t}+\frac{\partial }{\partial x_k}\left(v_k D_{ij}-v_j D_{ik}-\varepsilon_{jm k}\mathscr{E}_{B_{im}}\right)=-v_jg_i-J_{ij},\label{eul_final_d}\\[2mm]
&\displaystyle\frac{\partial f_i}{\partial t}+\frac{\partial f_i v_k}{\partial x_k}=0,\label{eul_final_e}\\[2mm]
&\displaystyle\frac{\partial g_i}{\partial t}+\frac{\partial (g_i v_k+J_{ik})}{\partial x_k}=0,\label{eul_final_f}
\end{align}
\end{subequations}
where   $J_{ij}=((\rho\mathscr{E})_{D_{ij}}-(\rho\mathscr{E})_{A_{ij}})/\rho$ and  $\Phi_{ij}=\mathscr{E}_{D_{ij}}$. The newly adopted state variable $\DD=[D_{ij}]$ has the physical interpretation of the rate of $\BB$ and  is connected with $\boldsymbol{g}$ by the relation ${\rm div}\DD=\boldsymbol{g}$. Note that (\ref{eul_final}) is a closed extension of (\ref{model_eul}) in the sense that no  new first order differential consequences like (\ref{eul_constr}) can be found.

Now we are in position to prove the energy conservation for (\ref{eul_final}). We assume that the total energy $\mathscr{E}$ does not depend on $\ff$ and $\boldsymbol{g}$, i.e.  $\mathscr{E}=\mathscr{E}(\vv, \AAA, \BB, \DD)$. In other words, the vectors $\ff$ and $\boldsymbol{g}$ play the role of auxiliary variables that allow to write equations (\ref{eul_final_c}) and (\ref{eul_final_d}) in a divergence (conservative) form. If we now multiply  Eqs.~(\ref{eul_final}) by the factors
\begin{equation}\label{mult_fin}
(\rho\mathscr E)_{\rho v_i},\ \  (\rho\mathscr E)_{A_{ik}},\ \  (\rho\mathscr E)_{B_{ij}},\ \  (\rho\mathscr E)_{D_{ij}},\ \ (\rho\mathscr E)_{f_{i}}\equiv 0,\ \ (\rho\mathscr E)_{g_{i}}\equiv0
\end{equation}
and sum all of them, we obtain
\begin{equation}\label{eul_en_fin}
\displaystyle\frac{\partial \rho \mathscr{ E}}{\partial t}+\frac{\partial }{\partial x_k}\left(v_k\rho \mathscr{E}+\varepsilon_{jmk}\mathscr{E}_{B_{ij}}\mathscr{E}_{D_{im}}
+\rho v_n (\rho \scE_\rho\delta_{nk} + A_{mk} \mathscr{E}_{A_{mn}} - B_{mk}\mathscr{E}_{B_{mn}}-D_{mk}\mathscr{E}_{D_{mn}})\right)=0.
\end{equation}
In order to get zero on the right hand side of (\ref{eul_en_fin}), we need to add  positive terms $c\mathscr{E}_{D_{ij}}\mathscr{E}_{D_{ij}}/\mathscr{E}_s\geq 0$ and $(1-c)\mathscr{E}_{D_{ij}}\mathscr{E}_{D_{ij}}/\mathscr{E}_s\geq 0$ to the entropy production terms $\varsigma_l$.

The total stress tensor $\TT=[T_{ik}]=\rho\FF\mathscr{U}^\mathsf{T}_\FF$ in (\ref{eul_final_a}) has the form
\[T_{ik}=-\rho(\rho\scE_\rho\delta_{ik} - B_{mk}\mathscr{E}_{B_{mi}} - D_{mk}\mathscr{E}_{D_{mi}} + A_{mk}\mathscr{E}_{A_{mi}}).\] This particular expression  for the stress tensor arises again (as we have it already seen in (\ref{piola}) and (\ref{total_stress_comp})) from the requirement of the energy conservation.

We now show that (\ref{eul_final}) does not possess the complete Godunov structure (\ref{xtcomp})  even though, as we have just seen,  it is the system of equations in the  conservative form that implies the energy conservation. Namely, equations (\ref{eul_final}) can not be symmetrized and, consequently, their mathematical regularity  remains an open problem.

To see that, we introduce the vector of conjugate variables $\pp$ composed of multipliers (\ref{mult_fin}), and let the potential $L(\pp)$ be the Legendre transformation of the potential $\rho\mathscr{E}$ with respect to the conservative variables $\qq^{(ext)}$.  Then (\ref{eul_final}) can be written as
\begin{subequations}\label{eulext}
\begin{align}
&\displaystyle\frac{\partial L_{v_i}}{\partial t}+\frac{\partial (M^k_{v_i}-L_{\beta_{mk}}\beta_{mi}-L_{\delta_{mk}}\delta_{mi}+ L_{\alpha_{mk}}\alpha_{mi})}{\partial x_k}=0,\label{eulext_a}\\[2mm]
&\displaystyle\frac{\partial  L_{\alpha_{ik}}}{\partial t}+\frac{\partial M^m_{\alpha_{im}}}{\partial x_k}=-\varepsilon_{kml}v_m L_{\beta_{il}}-\Delta_{ik},\\[2mm]
&\displaystyle\frac{\partial  L_{\beta_{ij}}}{\partial t}+\frac{\partial (M^k_{\beta_{ij}}-v_jL_{\beta_{ik}}+\varepsilon_{jm k}\delta_{im})}{\partial x_k}=-v_j f_i,\\[2mm]
&\displaystyle\frac{\partial L_{\Delta_{ij}}}{\partial t}+\frac{\partial (M^k_{\Delta_{ij}}-v_j L_{\Delta_{ik}}-\varepsilon_{jm k}\beta_{im})}{\partial x_k}=-v_jg_i-J_{ij},
\end{align}
\end{subequations}
where $M^k(\pp)=v_kL(\pp)$, $(\rho\mathscr{E})_{\rho v_i}=v_i$, $(\rho\mathscr{E})_{A_{ij}}=\alpha_{ij}$, $(\rho\mathscr{E})_{B_{ij}}=\beta_{ij}$, $(\rho\mathscr{E})_{D_{ij}}=\Delta_{ij}$, $J_{ij}=(\Delta_{ij}-\alpha_{ij})/\rho$. There is no need to introduce the potentials $M^k$. We  have  done it only in order to see more clearly the difference between (\ref{god_form1}) and (\ref{eulext}).
The difference from zero of the gauge constraints, in particular ${\rm div }\DD=\boldsymbol{g}\neq 0$, in the presence of dissipation is the main reason why (\ref{eulext}) can not be symmetrized in the manner of (\ref{eul_0}), (\ref{eul_sym}). Indeed, addition of  $\beta_{ij}f_i+\Delta_{ij}g_i \neq 0$ of the constraints ${\rm div}\BB=0$, ${\rm div}\DD=\boldsymbol{g}$ to (\ref{eulext_a}) violates the momentum conservation.

Finally, we emphasize that (\ref{eul_final}) in  the Lagrangian form  does possess  the Godunov structure. To see this, it is necessary to enlarged the original Lagrangian equations (\ref{model_lagr}) with the pair of complimentary equations (see the last remark in Section~\ref{sec:model_lagr}) for the new state variables $\BB$, $\DD$
\begin{subequations}
\begin{align*}
&\displaystyle\frac{{\rm d} \hat{B}_{im}}{{\rm d} t}+\varepsilon_{jnm}\frac{\partial \mathscr{U}_{\hat{D}_{in}} }{\partial y_j}=0,\\[2mm]
&\displaystyle\frac{{\rm d} \hat{D}_{im}}{{\rm d} t}-\varepsilon_{jnm}\frac{\partial \mathscr{U}_{\hat{B}_{in}} }{\partial y_j}=-\hat{J}_{im},
\end{align*}
\end{subequations}
or in the matrix notation
\begin{subequations}
\begin{align*}
&\displaystyle\frac{{\rm d} \hat{\BB}}{{\rm d} t}+{\rm rot}\mathscr{U}_{\hat{\DD}} =0,\\[2mm]
&\displaystyle\frac{{\rm d} \hat{\DD}}{{\rm d} t}-{\rm rot}\mathscr{U}_{\hat{\BB}}=-\hat{\JJ},
\end{align*}
\end{subequations}
where $\hat{\BB}=[\hat{B}_{ij}]=\FF^\mathsf{T}\BB$, $\hat{\DD}=[\hat{D}_{ij}]=\FF^\mathsf{T}\DD$, $\hat{\JJ}=[\hat{J}_{ij}]=\FF^\mathsf{T}\JJ$. The energy conservation for the extended Lagrangian model takes the form
\begin{equation*}\label{lagr_en_ext}
\frac{{\rm d}\mathscr{U}}{{\rm d}t}+\frac{\partial}{\partial y_j}\left(\mathscr{U}_c\mathscr{U}_{\hat{w}_j}-\mathscr{U}_{v_i}\mathscr{U}_{F_{ij}}+\varepsilon_{jnm}\mathscr{U}_{\hat{B}_{im}}\mathscr{U}_{\hat{D}_{in}}\right)=0.
\end{equation*}

Summing up, we have not succeeded to reformulate (\ref{model_eul}) completely into the form (\ref{xtcomp}) of Godunov's equations. We have however demonstrated that already the route leading to such reformulation brings interesting physical and mathematical insights about (\ref{model_eul}). This finding is in fact an indirect proof of the pertinence and importance (both from the physical and the mathematics-numerical point of view) of the Godunov structure.

We shall now leave the Eulerian framework and return for the rest of this paper to the one dimensional version of the Lagrangian framework.
\section{\label{sec:num}Numerical Illustrations}

In this section we turn our attention to  problems associated  with  finding numerical solutions to the time evolution equations introduced in Section \ref{sec:model_lagr}. In this paper we restrict ourself by considering the system (\ref{model_lagr}) in one-dimensional space and with no dissipation, i.e. with  all the source terms in (\ref{model_lagr}) equal zero. This means that the numerical illustrations below provide just a basic structure of solutions of Eqs.~(\ref{model_lagr})  and give us an opportunity to show how is the Godunov structure directly used in numerical calculations. In order to  explain the first statement, it suffices to recall \cite{toro}  that in the theory of nonlinear hyperbolic partial differential equations,  solutions to the Riemann problem (the initial value problem with piecewise constant initial data) provide an information about characteristics which then provides the basic framework for all solutions.

Our general strategy with which we approach  numerical calculations is an attempt to regard the modifications of  continuum formulations  needed in such  calculations (in particular the discretization) as  a physically meaningful reduction to  more macroscopic levels of description. This means in particular that we shall try to preserve the mathematical structure of the continuum formulation (expressing, as we have seen, the compatibility with mechanics and thermodynamics) in the discrete formulation. This type of physically meaningful discretization has been introduced by Godunov \cite{god1959}  for the system of ideal hydrodynamics. Our objective in this section is to present the method, known as  the Godunov numerical scheme, and illustrate it in the context of the governing equations derived in  previous sections.

The point of departure of the Godunov numerical scheme is  a system of hyperbolic conservation laws. But (\ref{model_eul}) does not belong to such class. We have  only started to transform it into this form in Section \ref{eulc}. On the other hand, the Lagrangian governing equations (\ref{model_lagr}) do belong to the Godunov class and thus to the class of hyperbolic conservation laws but, as it was explained in Section \ref{sec:lagr_limit}, they are not well suited to deal with both solid deformations and liquid flows except when we restrict ourselves to one Lagrangian dimension. We shall therefore turn now to this special case (see the next section for a precise definition of fluids in one Lagrangian dimension)  even if such fluids obviously do not reflect the full complexity of  real yield-stress fluids.
Nevertheless, this (toy) example gives us a possibility to present the Godunov numerical scheme and at the same time to explore a basic framework  for  solutions of the solid-fluid mixture model introduced in this paper.

\subsection{\label{god_method}Godunov numerical method}

The system of equations (\ref{model_lagr}) can be conveniently written in one dimension ($y_1=y$) and without the source terms as one vectorial conservation law
\begin{equation}\label{num_1d}
\frac{{\rm d} \qq}{{\rm d} t}+\frac{\partial \boldsymbol{\mathcal{F}}(\qq)}{\partial y}=0,
\end{equation}
where
\[\qq=(v_1,v_2,v_3,F_{11},F_{21},F_{31},c,\hat{w}_1)^{\mathsf{T}}\]
is the vector of conserved variables and
\[
\boldsymbol{\mathcal{F}}(\qq)=(-\mathscr{U}_{F_{11}},-\mathscr{U}_{F_{21}},-\mathscr{U}_{F_{31}},-\mathscr{U}_{v_1},-\mathscr{U}_{v_2},-\mathscr{U}_{v_3},\mathscr{U}_{\hat{w}_1},\mathscr{U}_{c})^{\mathsf{T}}
\] is the vector of fluxes. It is important to note that even if we restrict ourselves only to one Lagrangian coordinate $y_1$, we still have a fluid with three components of velocities $\vv$ and three components (first column) of the strain tensor $\FF$. For the sake of brevity, we further  omit the subscript ``1'' and  write  hereafter $\qq=(v_1,v_2,v_3,F_{1},F_{2},F_{3},c,\hat{w})^{\mathsf{T}}$.

Following to Refs.~\cite{god1959,toro,leveque}, we discretize (\ref{num_1d})  by using  the first order Godunov method.
The discrete form of \eqref{num_1d} in a control volume $[y_m, y_{m+1}] \times [t^n, t^{n+1}]$ of dimensions $\Delta y=y_{m+1}-y_m$, $\Delta t=t^{n+1}-t^n$ becomes
\[\qq^{n+1}_{m+1/2} = \qq^{n}_{m+1/2}-\frac{\Delta t}{\Delta x}(\boldsymbol{\mathsf{F}}_{m+1}-\boldsymbol{\mathsf{F}}_m).\]
The quantity  $q^{n+1}_{m+1/2}$ approximates the average value of $\qq$ in the $m$th interval at time $t^n$:
\[\qq^{n+1}_{m+1/2}\approx \frac{1}{\Delta y}\int_{y_m}^{y_{m+1}}\qq(t_n,y)dy,\]
and $\boldsymbol{\mathsf{F}}_{m}$ is an approximation to the average flux along $y=y_{m}$:
\[\boldsymbol{\mathsf{F}}_{m} \approx \frac{1}{\Delta t}\int_{t_n}^{t_{n+1}}\boldsymbol{\mathcal{F}}(\qq(t,y_{m}))d t.\]

In Godunov type methods, $\boldsymbol{\mathsf{F}}_m=\boldsymbol{\mathcal{F}}(\QQ_m)$, where $\QQ_m=\QQ_m(\qq^n_{m-1/2},\qq^n_{m+1/2})$ is the two-argument vector valued function that is obtained as a solution (exact or approximate) of the local Riemann problem, i.e as a solution of the initial value problem with the piecewise initial data:
\[
\qq(t^n,y)=\left\{\begin{array}{l}
\qq^n_{m-1/2},\ \ y<y_m,\\
\qq^n_{m+1/2},\ \ y>y_m.
\end{array}\right.\]
The Riemann problem is solved in this paper by using an approximate method based upon the characteristic tracing. We therefore need to have a detailed knowledge of the eigenvalues and the eigenvectors of~(\ref{num_1d}).

By introducing
$\pp=(\mathscr{U}_{v_1},\mathscr{U}_{v_2},\mathscr{U}_{v_3}, \mathscr{U}_{F_{1}},\mathscr{U}_{F_{2}},\mathscr{U}_{F_{3}},\mathscr{U}_{c},\mathscr{U}_{\hat{w}})^{\mathsf{T}}$, also called primitive variables, Eqs. (\ref{num_1d}) can be rewritten as a symmetric quasi-linear system (\ref{eq:god_sym}):
\begin{equation}\label{num_sym}
\mathsf{A}(\pp)\frac{{\rm d} \pp}{{\rm d}t}+\mathsf{B}(\pp)\frac{\partial \pp}{\partial y}=0.
\end{equation}
The $8\times8$-matrices $\mathsf{A}$, $\mathsf{B}$ appearing in this system are given by
\begin{eqnarray}\label{matrixA}
\mathsf{A}=\left[\begin{array}{cccccccc}
 1 & 0 & 0 & 0 & 0 & 0 & 0 & 0 \\
 0 & 1 & 0 & 0 & 0 & 0 & 0 & 0 \\
 0 & 0 & 1 & 0 & 0 & 0 & 0 & 0 \\
 0 & 0 & 0 & \mathscr{U}_{F_1F_1} & \mathscr{U}_{F_1F_2} & \mathscr{U}_{F_1F_3} & \mathscr{U}_{F_1c} & \mathscr{U}_{F_1\hat{w}} \\
 0 & 0 & 0 & \mathscr{U}_{F_1F_2} & \mathscr{U}_{F_2F_2} & \mathscr{U}_{F_2F_3} & \mathscr{U}_{F_2c} & \mathscr{U}_{F_2\hat{w}} \\
 0 & 0 & 0 & \mathscr{U}_{F_1F_3} & \mathscr{U}_{F_2F_3} & \mathscr{U}_{F_3F_3} & \mathscr{U}_{F_3c} & \mathscr{U}_{F_3\hat{w}} \\
 0 & 0 & 0 & \mathscr{U}_{F_1c}   & \mathscr{U}_{F_2c} & \mathscr{U}_{F_3c} & \mathscr{U}_{cc} & \mathscr{U}_{c\hat{w}} \\
 0 & 0 & 0 & \mathscr{U}_{F_1\hat{w}}   & \mathscr{U}_{F_2\hat{w}} & \mathscr{U}_{F_3\hat{w}} & \mathscr{U}_{c\hat{w}} & \mathscr{U}_{\hat{w}\hat{w}} \\
\end{array}\right]^{-1},
\end{eqnarray}
\[
\mathsf{B}=\left[\begin{array}{cccccccc}
 0 & 0 & 0 & 1 & 0 & 0 & 0 & 0 \\
 0 & 0 & 0 & 0 & 1 & 0 & 0 & 0 \\
 0 & 0 & 0 & 0 & 0 & 1 & 0 & 0 \\
 1 & 0 & 0 & 0 & 0 & 0 & 0 & 0 \\
 0 & 1 & 0 & 0 & 0 & 0 & 0 & 0 \\
 0 & 0 & 1 & 0 & 0 & 0 & 0 & 0 \\
 0 & 0 & 0 & 0 & 0 & 0 & 0 & -1 \\
 0 & 0 & 0 & 0 & 0 & 0 & -1 & 0 \\
\end{array}\right] \ =\ \mbox{constant}.\nonumber
\]
In Appendix \ref{ap:matrix} we give formulas  for the entries  of the matrix $\mathsf{A}$.

A natural way to define an approximate Riemann solution is to replace nonlinear problem (\ref{num_sym}) by a linearized problem
\[
\mathsf{A}_m\frac{{\rm d} \pp}{{\rm d}t}+\mathsf{B}\frac{\partial \pp}{\partial y}=0
\]
that is defined locally at each cell interface $y_m$. In this paper, we employ the basic approximation $\mathsf{A}_m=\mathsf{A}((\pp_{m-1/2} + \pp_{m+1/2})/2)$. The linearized problem can be now rewritten in a characteristic form
\[\frac{{\rm d}\cc}{{\rm d} t}+\mathsf{S}_m\frac{\partial \cc}{\partial y}=0\]
inside of each two cells $[y_{m-1},y_m]$, $[y_m,y_{m+1}]$. Here, $\mathsf{S}_m={\rm diag}(s_1^m,s_2^m,\ldots,s_8^m)$ is a diagonal matrix with local sound velocities on the diagonal, $\cc$ is the vector of characteristic variables. Matrix $\mathsf{S}_m$ and vector $\cc$ can be computed using the formulae
\begin{equation}\label{num_eig}
\mathsf{S}_m = \mathsf{R}_m^{\mathsf{T}}\mathsf{C}_m\mathsf{R}_m,\ \ \  \cc=\mathsf{R}^\mathsf{T}_m\mathsf{A}^{\frac{1}{2}}_m \pp,\ \ \ \mathsf{R}_m^\mathsf{T}\mathsf{R}_m=\mathsf{I},
\end{equation}
where $\mathsf{R}_m$ is the matrix of eigenvectors of the symmetric matrix
$\mathsf{C}_m=\mathsf{A}_m^{-\frac{1}{2}}\mathsf{B}\mathsf{A}_m^{-\frac{1}{2}}$ and $\mathsf{I}$ is the $8\times 8$-identity matrix. Recall  that the mixture total energy $\mathscr{U}$ is supposed to be a convex function, consequently the matrix $\mathsf{A}_m$ is positive definite and the matrices $\mathsf{A}_m^{-\frac{1}{2}}$ and  $\mathsf{A}_m^{\frac{1}{2}}$  are exist. With the formulae (\ref{num_eig}) it is easy to obtain the solution $\PP_m=\PP_m(\pp_{m-1/2},\pp_{m+1/2})$ of the Riemann problem on each cell interface $y=y_m$ (see for example Refs.~\cite{toro,leveque}).

Unfortunately, in general, the structure of the matrix $\mathsf{A}_m$ is dense and it is impossible to obtain  exact expressions for $\mathsf{A}_m^{\frac{1}{2}}$, $\mathsf{C}_m$ and $\mathsf{R}_m$. However, since $\mathsf{A}_m$ and $\mathsf{C}_m$ are symmetric matrices,  it is possible to use modern fast algorithms of numerical linear algebra \cite{lapack}.

Before leaving this section we make  an observation about  sound velocities $s_1,...,s_8$ (we assume that they are arranged in  ascending order $s_1\leq s_2\leq\ldots\leq s_8$). The following three statements are true (see also Ref.~\cite{rom2013}):
\begin{itemize}
\item[1.]
$s_m\leq 0$, $m=1,2,3,4$ and $s_m\geq 0$, $m=5,6,7,8$.

\item[2.]

Sound velocities converge  to the next values
\begin{eqnarray*}
|s_1|=s_8\rightarrow c_l, \ \ \ |s_2|=s_7\rightarrow b_0,\ \ \ |s_3|=s_6=|s_4|=s_5\rightarrow c_t
\end{eqnarray*}
in the limit of the pure solid (i.e. when $c\rightarrow 1$). The quantities appearing in the previous line have been introduced in (\ref{eos_solid}) and (\ref{eos_fluid}). In the pure liquid limit (i.e. when $c\rightarrow 0$) at rest  the sound velocities of the mixture become
\begin{eqnarray*}
|s_1|=s_8\rightarrow d_0, \ \ |s_2|=s_7=\rightarrow b_0,\ \ \ |s_3|=s_6=|s_4|=s_5\rightarrow 0.
\end{eqnarray*}

\item[3.]
If $\ww \neq (0,0,0)$ then, in general, the distribution of the velocities $s_m$ relative to zero is not symmetric:
\[|s_1|\neq|s_8|,\ \ |s_2|\neq|s_7|,\ \ |s_3|\neq|s_6|,\ \ |s_4|\neq|s_5|.\]
\end{itemize}

\subsection{\label{NT}Numerical tests}

In this section we  work out several test Riemann problems for solid-fluid mixtures. We recall that an investigation of solutions to the Riemann problem is one of the  standard ways to investigate properties of solutions to a system of nonlinear hyperbolic  partial differential equations.

In all three tests  we separate the fluid under investigation into two sections (we call them left and right sections)  and let the two sections  to collide. At the point of separation we thus have initially a discontinuity in velocities, otherwise all other properties change continuously.

In the  tests we assume that:

\begin{itemize}

\item[1.]
The mixture has constant volume fractions $\alpha$, namely
0.6 for solid  and  0.4 for liquid. The computational domain
is the  interval $[-0.5; 0.5]$ (in centimeters). In each test
the initial data defines a Riemann problem with a left
section $[-0.5; 0)$ and a right section $(0;0.5]$. The material parameters are chosen to be the following: $\rho_1=\rho_2=\rho=1$ ${\rm g/cm^3}$, $c_l=1.5$ ${\rm km/s}$, $c_t=0.5$ ${\rm km/s}$, $b_0=d_0=1.38$ ${\rm km/s}$, $\gamma=3,$ $c_{V}=1$.
The ``prepared'' mixture at the reference stress-free configuration has the next sound velocities (km/s):
\[
\begin{array}{c}
-s_1=s_8=1.513,\ \ \  -s_2=s_7=0.818,\\[4mm]
-s_3=-s_4=s_5=s_6=0.387.
\end{array}
\]

\item[2.]
The CFL coefficient  is put to be equal to 0.9. All figures (except in the shear test) represent a numerical solution in the Eulerian frame, i.e. the Lagrangian computational mesh is moving at the mixture velocity.

\item[3.]
The temperatures of both phases are the same. We are making this simplifying assumption for the following reason.
In order to study  discontinuous solutions, the  two equations for phase entropies in (\ref{model_lagr}) should be replaced by two relations which describe correctly the Rankine-Hugoniot shock conditions. In the case of a  single fluid (or a single solid), the energy conservation law must replace  the entropy conservation law. In our case,  we have only one energy conservation law for the mixture (\ref{lagr_en}). Hence, an extra jump condition (for example for one of the phase energies) is needed for formulating the correct  Rankine-Hugoniot relations.
In order to  overcome this difficulty we follow  Ref.~\cite{rom_drik2010} and restrict ourselves to a special case in which  the temperatures of both phases are assumed to be  equal (i.e. $T_l=\partial U/\partial
S_l=\partial U_l/\partial s_l$, $l=1,2$ are equal). From the physical point of view this means that we assume that the relaxation leading to the equilibration of the two temperatures proceeds much faster than the rest of the time evolution.
The reduced system of
governing equations corresponding to  the case $T_1=T_2$ can be derived from (\ref{model_lagr}) by assuming
that thermal effects can be characterized by a single  mixture entropy $S = c s_1 + (1-c) s_2$. By solving the
system of equations \[T_1=\frac{\partial U_1}{\partial s_1}=\frac{\partial U_2}{\partial s_2}=T_2, \ \ c s_1+(1-c)s_2=S.\] we then obtain the
phase entropies $s_l$ as   functions of the volume and the mass fractions, the phase densities and the
mixture entropy: $s_l(c,\alpha,\rho_1,\rho_2,S)$, $l=1,2$.
The simplified system of governing equations representing the special case in which $T_1=T_2$ can be now derived from
(\ref{model_lagr}) by replacing the two equations for $S_1$, $S_2$ by a single energy conservation law (\ref{lagr_en}).

\end{itemize}

\subsubsection*{Test 1: Longitudinal perturbation; symmetric collision }

The initial conditions in the first test are:
\begin{itemize}
\item
$v_1=1\ {\rm km/s}$ for the left section;
\item
$v_1=-1\ {\rm km/s}$ for the right section
\end{itemize}
and $F_{1}=1$, $F_{2}=F_{3}=0$, $c=0.6$, $\hat{w}=0$ km/s,
for the both sections.

\begin{figure}
\includegraphics[trim = 30mm 70mm 20mm 10mm scale=100.0]{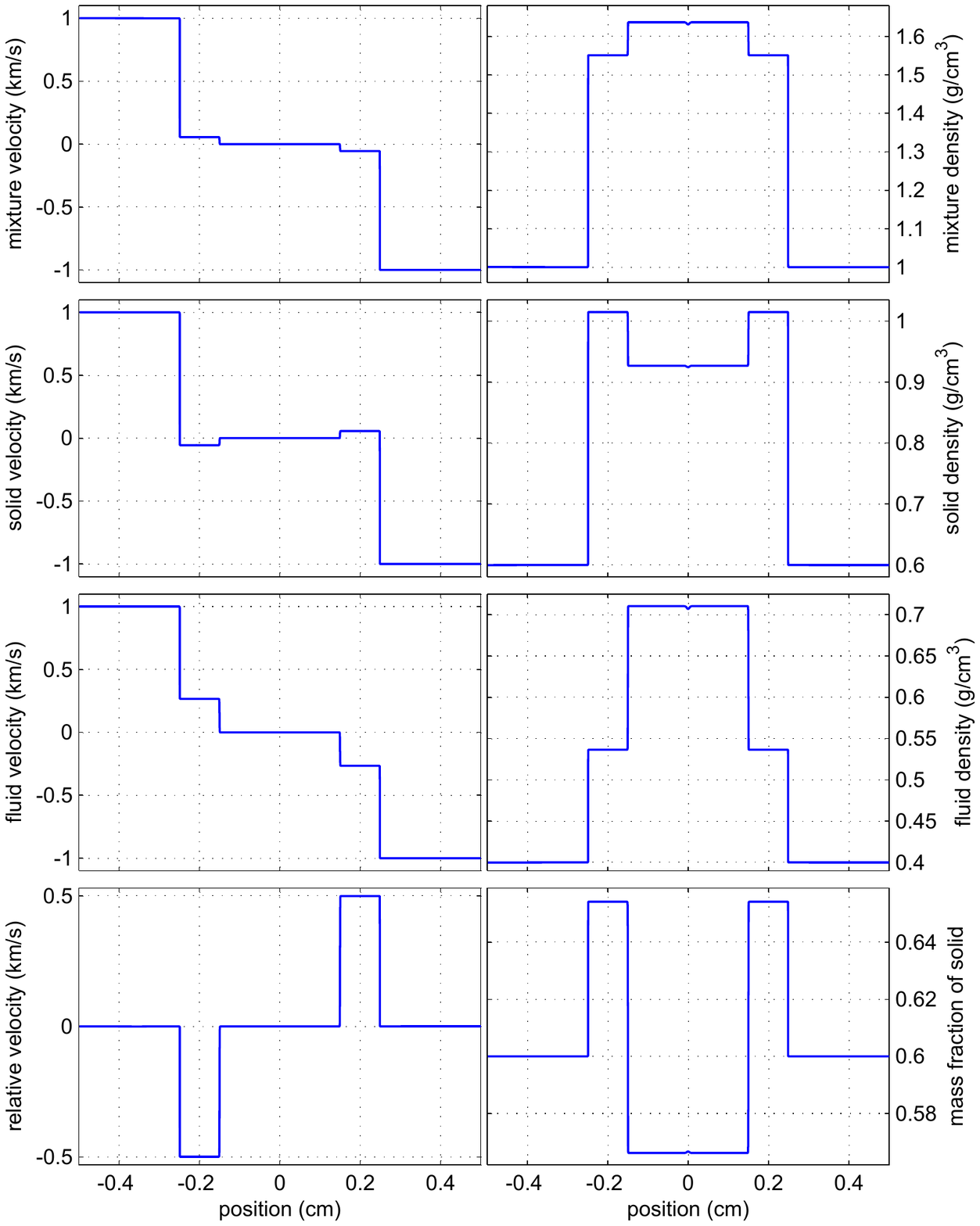}
\caption{\it Symmetric collision; numerical solution at the time $t=0.15\times10^{-5}$ s}
\label{fig:shock1}
\end{figure}

Fig. \ref{fig:shock1} shows the numerical solution at time $t=0.15\times10^{-5}$~s computed on a fine mesh of 12800 cells. The initial discontinuity in the mixture velocity breaks up into the four discontinuous waves of two types propagating out of the place of the initial discontinuity ($y=0$).

The two waves of the first type (fast) represent shocks which compress the mixture and the both phases. They propagate at the supersonic speed $\approx 2.66$ km/s.

The other two waves represent a second type (slow). They look also as discontinuities but, in contrast to the fast waves, they also compress the entire mixture and the fluid phase but density of the solid phase decreases. These waves propagate at the subsonic speed $\approx 1.055$ km/s while the mixture sound velocities after compression of the material in the first waves become as follow (km/s): $|s_1|=6.210$, $s_8=6.345$, $|s_2|=3.130$, $s_7=2.993$, $|s_3|=|s_4|=s_5=s_6=0.420$.

\begin{figure}
\includegraphics[trim = 30mm 170mm 20mm 35mm scale=0.15]{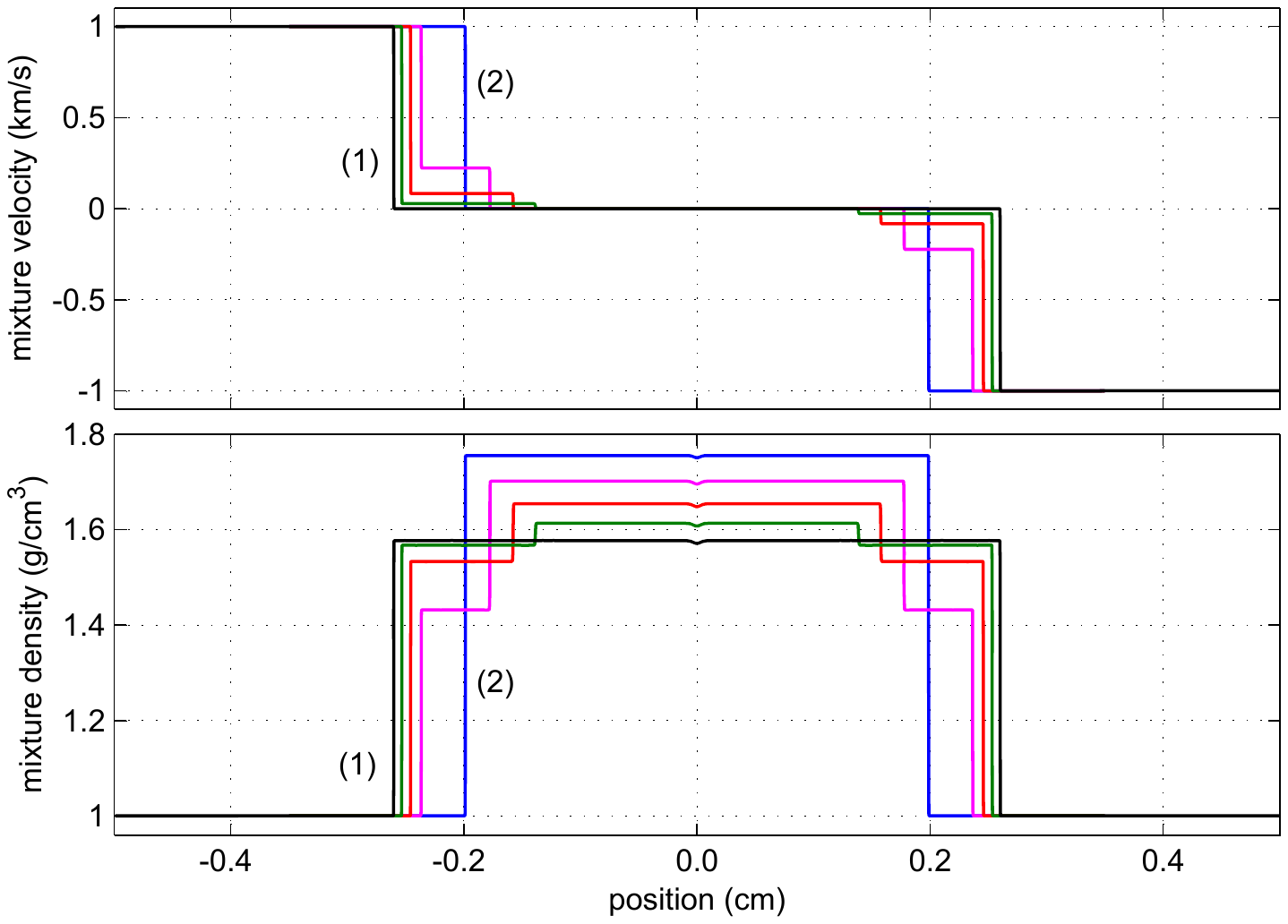}
\caption{\it (Colour online) Symmetric collision;  numerical solution at the time $t=0.15\times10^{-5}$~s for the different volume fractions. The   curve (1) corresponds to $\alpha=1-10^{-7}$ solid limit (black line), the curve (2) corresponds to $\alpha=10^{-7}$ fluid limit (blue line), the curves between (1) and (2) correspond to $\alpha=0.75$ (green line), $\alpha=0.5$ (red line), $\alpha=0.25$ (pink line).}
\label{fig:shock2}
\end{figure}

Fig. \ref{fig:shock2} depicts numerical solutions for different volume fraction $\alpha$ that varies from solid $\alpha\approx 1$ to fluid $\alpha\approx 0$ limit. The initial data are the same as in Fig. \ref{fig:shock1}. Curves 1 ($\alpha=1-10^{-7}$) and 2 ($\alpha=10^{-7}$) on Fig. \ref{fig:shock2} correspond to the case of the pure solid and of the pure fluid  respectively. It is important to remark that these curves are in excellent agreement with the results of computations for the single component equations with given equations of state (\ref{eos_solid}) and (\ref{eos_fluid}).

\subsubsection*{Test 2: Longitudinal perturbation; symmetric rarefaction }

The second test addresses  an expansion with the following discontinuous initial data:
\begin{itemize}
\item
$v_1=-0.25\ {\rm km/s}$ for the left section;
\item
$v_1= 0.25\ {\rm km/s}$ for the right section
\end{itemize}
and $F_{1}=1$, $F_{2}=F_{3}=0$, $c=0.6$, $\hat{w}=0$ km/s
for  both sections.

\begin{figure}
\includegraphics[trim = 30mm 70mm 20mm 20mm scale=0.5]{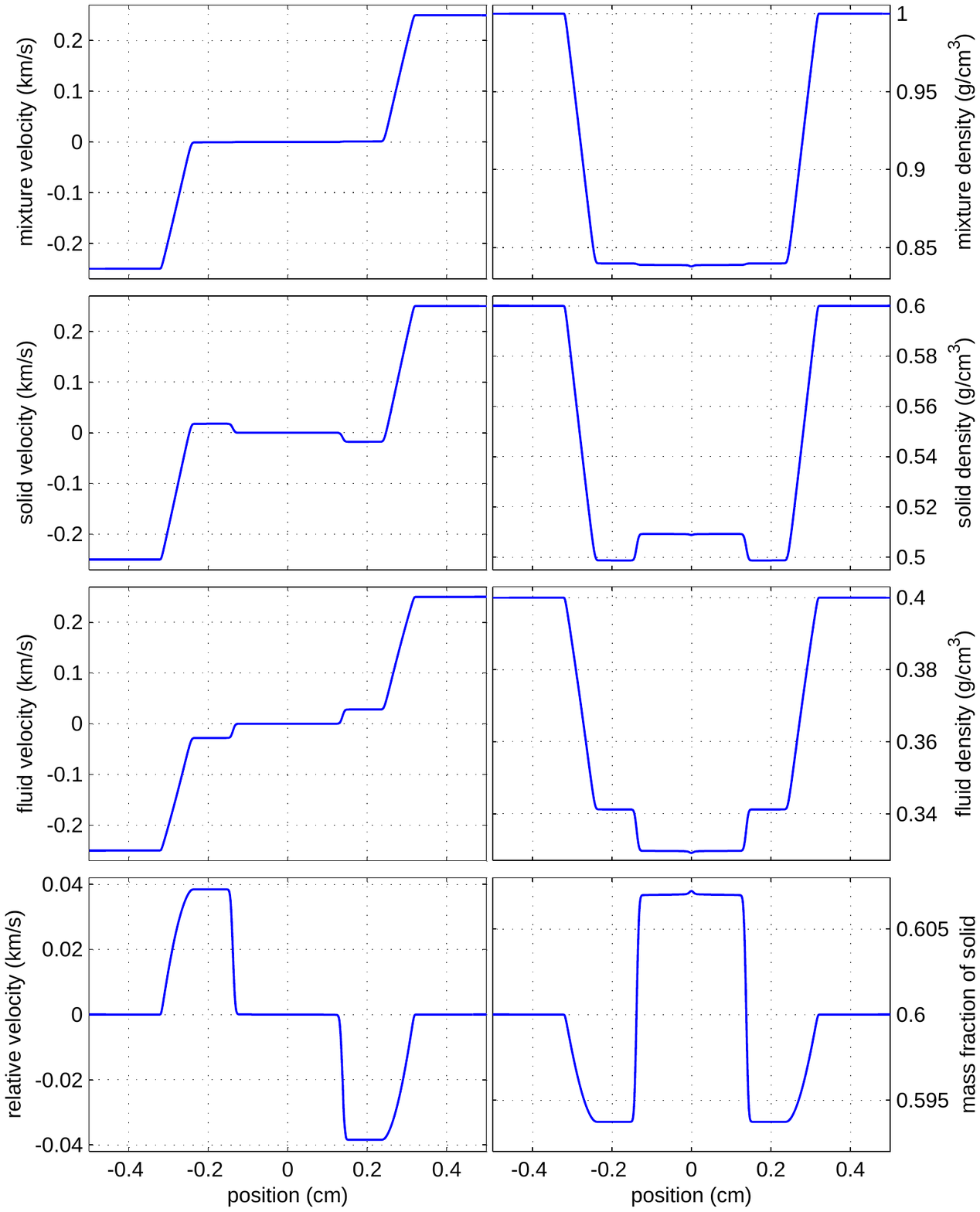}
\caption{\it Symmetric rarefaction;  numerical solution at the time $t=0.18\times10^{-5}$ s.}
\label{fig:rare1}
\end{figure}

Fig. \ref{fig:rare1} shows the numerical solution at time $t=0.18\times10^{-5}$~s computed with the  mesh of 12800 cells. The initial discontinuity in the  velocity breaks up into the four weak discontinuous waves (discontinuity in solution's derivatives) propagating out of the place of the initial discontinuity ($y=0$). These waves are also split  into two types.

The  waves of the first type (fast waves) look like pure rarefaction waves since  the densities of the mixture and the densities of both phases decrease. The character of waves of the second type (slow waves) is more complicated. The situation is similar  to the shock test. The density of the mixture and fluid phase  decreases but the solid phase is compressed in these waves. The front points of the fast waves  propagate at the transonic speed $\approx 1.52$~km/s.

\begin{figure}
\includegraphics[trim = 30mm 170mm 20mm 35mm scale=0.5]{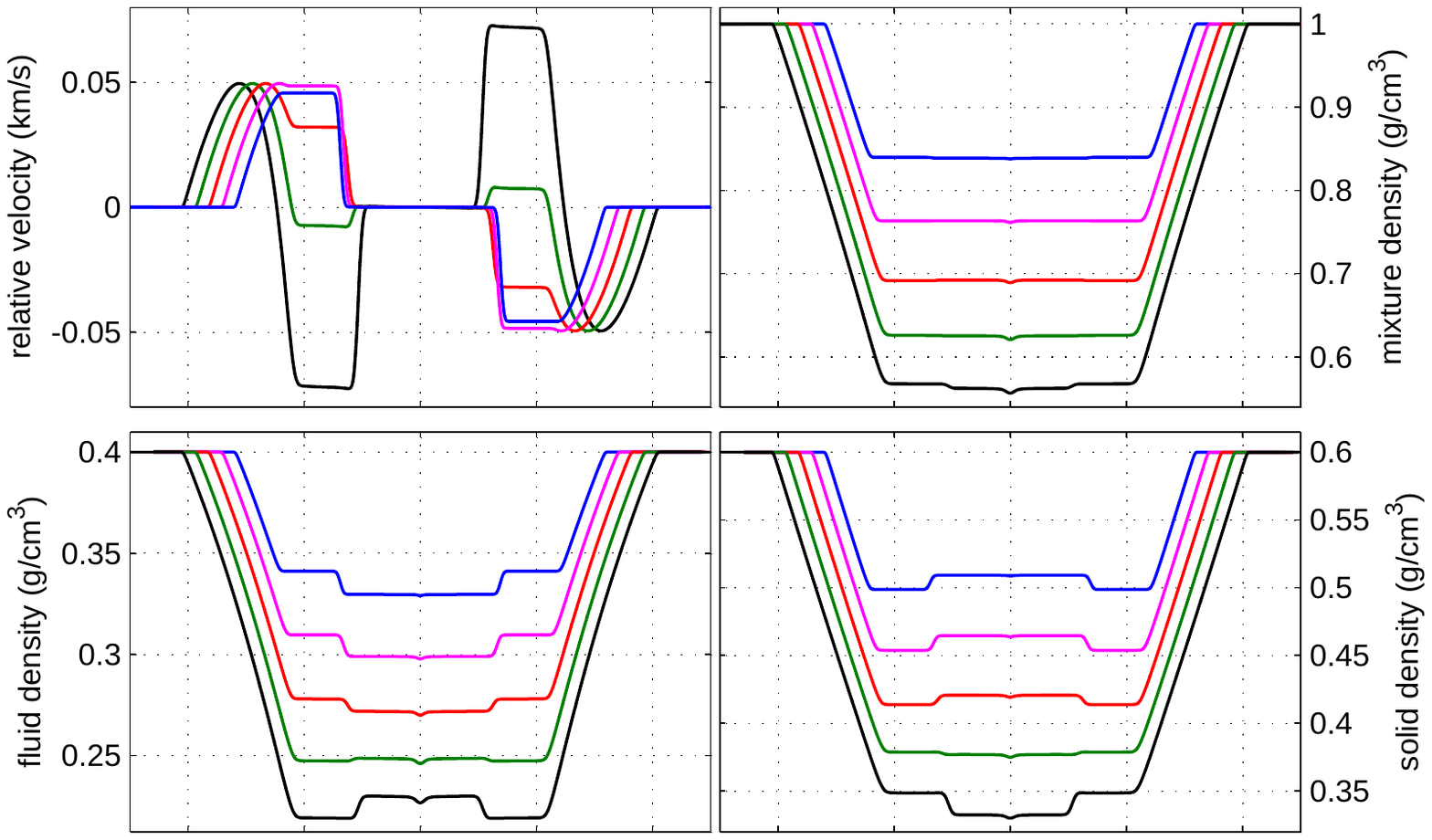}
\caption{\it (Colour online) Symmetric rarefaction; numerical solution at the time $t=0.18\times10^{-5}$ s for different initial data. The absolute value of the velocity with which the two parts move apart from each other is 0.25~km/s (blue curve); 0.375~km/s (pink curve);  0.5~km/s (red curve); 0.625~km/s (green curve);  0.75~km/s (black curve)}
\label{fig:rare2}
\end{figure}

Note that the behaviour of the two phases can change in  slow waves. Fig. \ref{fig:rare2} depicts a  comparison of the numerical solution profiles for different values of the expansion velocity. We see  that the phase behavior changes  from compression to expansion  for the  solid phase and in the reverse order  for the fluid phase. No such changes are observed for the shock test over the range of the collision velocities from $0.2$ km/s to $10$ km/s.

\subsubsection*{Test 3: Transversal perturbation; symmetric shear }

The third test  addresses   a  behavior  that is more complex than the one seen in the two previous tests. We consider  a shear displacement  with discontinuity in the second component of the velocity. Initial data are  the following:
\begin{itemize}
\item
$v_2=0.5\ {\rm km/s}$ for the left section;
\item
$v_2=-0.5\ {\rm km/s}$ for the right section
\end{itemize}
and $F_{1}=1$, $F_{2}=F_{3}=0$, $c=0.6$, $\hat{w}=0$ km/s
for the both sections. Note  that in the one dimensional case for the $y_1$ direction we have ${\rm d}\hat{w}_j/{\rm d}t=0$, ${\rm d}F_{ij}/{\rm d}t=0$,  $j=2,3$, and consequently  $w_2=w_3=0$ and shear phase velocity are equal.

\begin{figure}
\includegraphics[trim = 30mm 70mm 20mm 20mm scale=0.25]{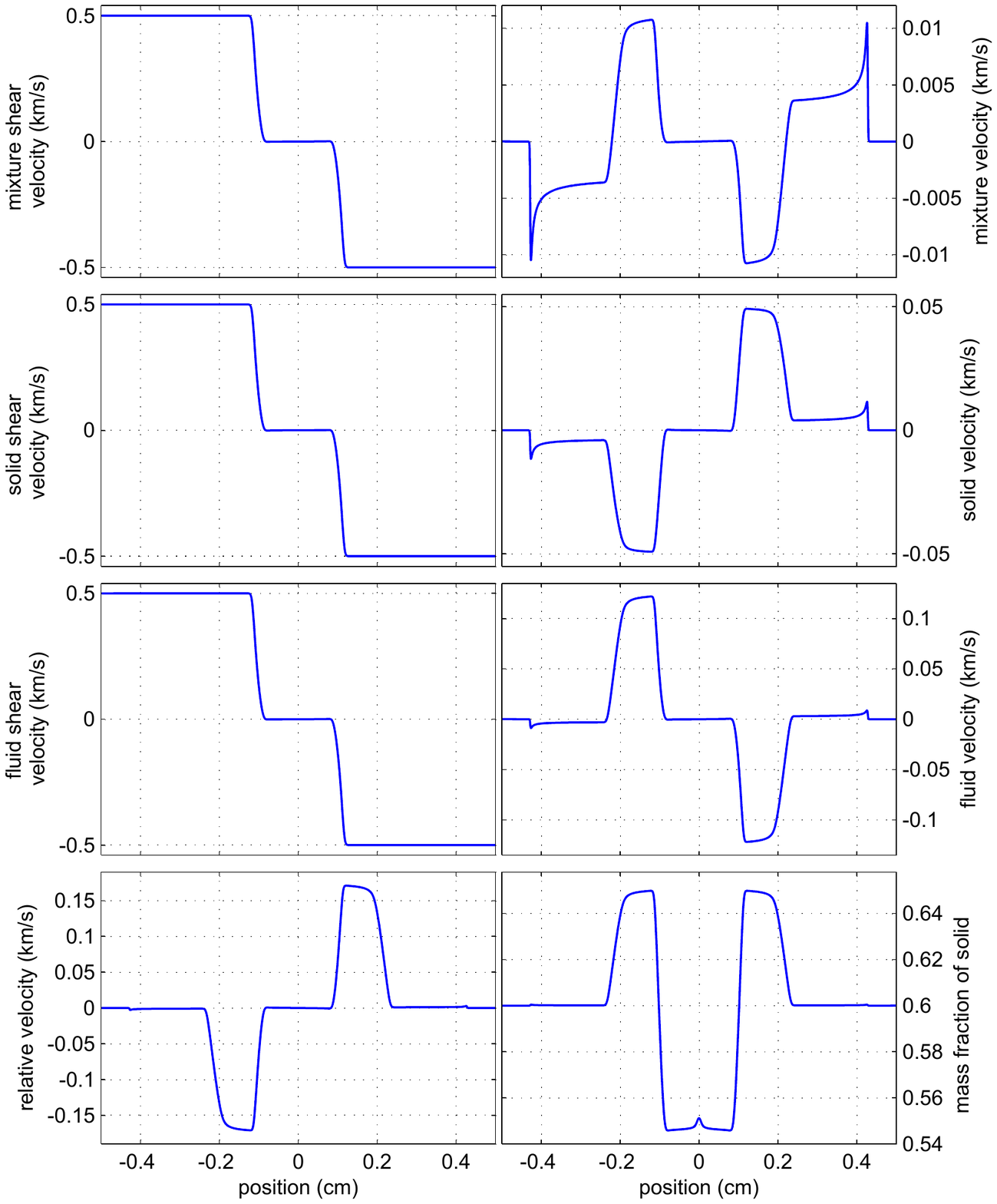}
\caption{\it Symmetric shear;  numerical solution at the time $t=0.28\times10^{-5}$ s.}
\label{fig:shear1}
\end{figure}

\begin{figure}
\includegraphics[trim = 30mm 110mm 20mm 20mm scale=0.5]{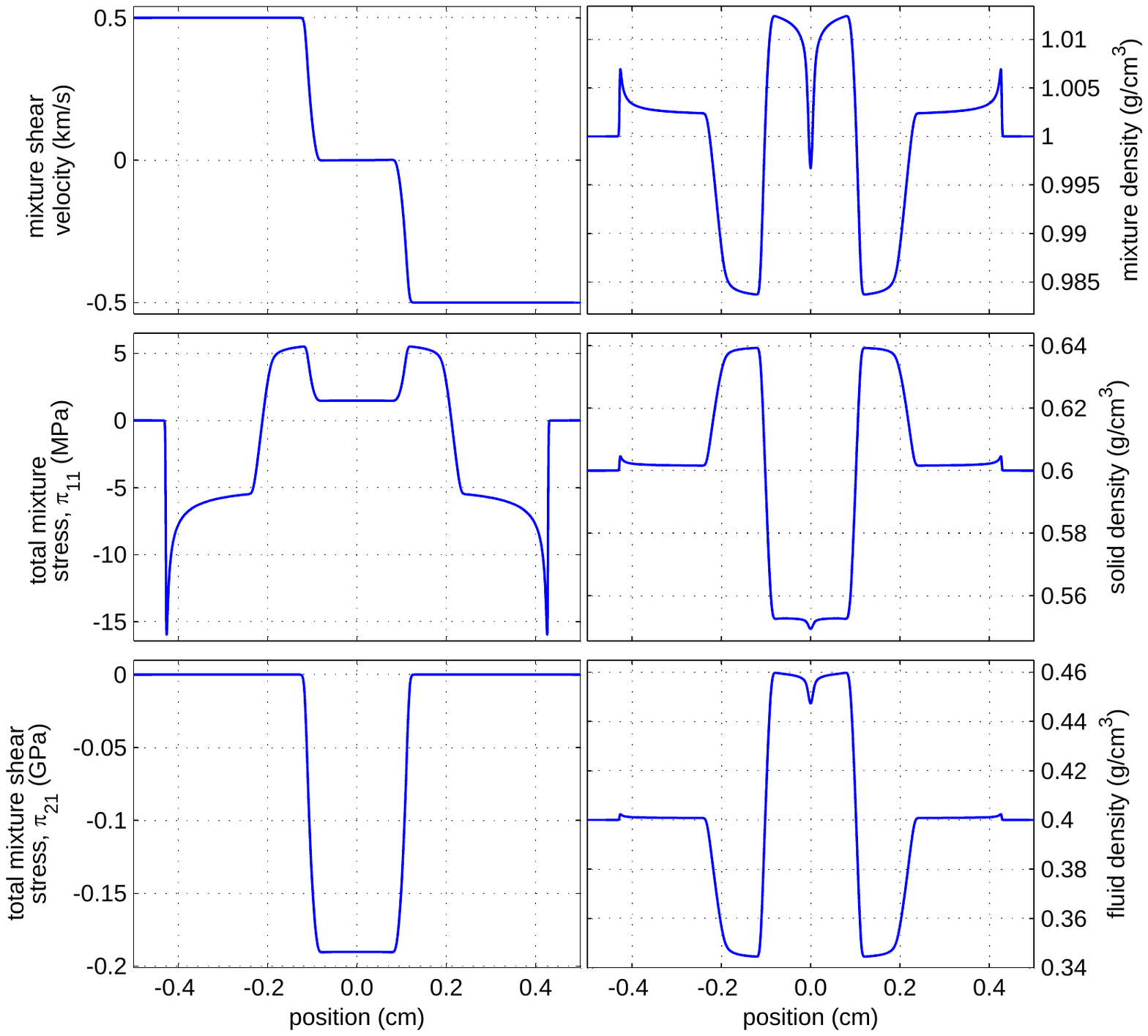}
\caption{\it Symmetric shear;  numerical solution at the time $t=0.28\times10^{-5}$ s.}
\label{fig:shear2}
\end{figure}

Fig. \ref{fig:shear1} and Fig. \ref{fig:shear2} show the numerical solutions at the time $t=0.28\times10^{-5}$ sec computed with the  mesh of 12800 cells. The main difference between the  shear test problem and the  two previous tests is that perturbations in the transversal direction ($y_2$ or $y_3$) lead to an appearance  of small perturbations propagating in the longitudinal direction $y=y_1$ (see for example the mixture velocity or total mixture stress $U_{F_{11}}$). The two fastest longitudinal waves have a small amplitude and look as  shocks since they have discontinuous profiles and they compress the mixture as well as   both phases. These waves propagate at the transonic speed $\approx 1.53$ km/s. There are also at least six longitudinal waves between two fastest longitudinal waves.

Finally, we note that the transversal waves are only of  one type (see mixture shear velocity, phase shear velocities and shear stress). In this example, they propagate at the sound shear velocity $\approx 0.384$ km/s.

\section{Concluding Remarks}

The physical systems under investigation in this paper are  solid-fluid mixtures. Their morphology is characterized by the mass and volume fractions, strain tensor, and relative velocity of the two phases. The time evolution equations are formulated in the top-down manner. The point of departure is the requirement of  compatibility of the time evolution with mechanics and thermodynamics. Such requirement  can be mathematically expressed in either Clebsch structure (that has emerged in carrying   the Hamiltonian formulation of Euler's equations toward thermodynamics) or the Euler structure (that has emerged in carrying the Euler local conservation laws toward thermodynamics).

In this paper we construct the governing equations as particular realizations of the Euler structure (more specifically its subclass known as the Godunov structure). In order to  bring
an additional insight into the physical and the mathematical content of the structure and of its particular realizations, we present both the Clebsch and the Godunov structures and compare them.
The time evolution equations are formulated in the paper
in both Lagrangian and Eulerian frameworks. The formulation in the Lagrangian framework is shown to possess the complete Godunov structure. Attempts to  bring also the formulation in the Eulerian framework to the complete Godunov form lead to the emergence of gauge constraints and a need for extensions. We intend to continue this line of research, offering new physical and mathematical insights into the Eulerian formulation,  in our future work.

As for the applications, we have in mind in this paper mainly  yield-stress fluids (our governing equations in the Eulerian framework can be  seen for instance as an extension of the  model of yield-stress fluids formulated recently in Ref.~\cite{burg1}). Other possible  applications may include for example   problems involving diffusion inside of elastoplastic solids \cite{rom2013,cald,grm2002} or diffuse interfaces~\cite{favr2009}.

The  problem of finding numerical solutions to the governing equations is also
approached with  physics in mind. Our intention is to preserve in the discretization the mathematical structure (expressing mathematically the physics, namely  the compatibility of the time evolution with mechanics and thermodynamics)  of the continuum formulation.  For partial differential equations belonging to the Godunov class of conservation laws,  the numerical method satisfying this requirement is the Godunov numerical scheme. The only case in which the governing equations introduced in this paper are both suitable for dealing  with  solid deformations and fluid flow,  and  possess the complete Godunov structure,  is the case of the Lagrangian equations in one Lagrangian dimension. We therefore limit in this paper our numerical illustrations to this special case.

By using the Godunov numerical scheme, we have worked out three tests. In all three tests we assume  the volume fraction to be fixed and we also assume the absence of the source terms causing dissipation. The unknown fields (functions of one Lagrangian coordinate) are three components of the velocity of the mixture, three components of the strain, mass fraction, and one component of the Lagrangian relative velocity. The initial condition has in all three tests a discontinuity in one of the components of the velocity of the mixture. From the physical point of view, we thus investigate evolution of the two-phase mixture that follows a collision of two sections of the mixture.

\section{Acknowledgments}
This research was partially supported by the Natural Sciences and Engineering
Research Council of Canada. The third author was partially supported by
the Russian Foundation for Basic Research (grant 13-05-12051), the Siberian Branch of Russian Academy of Sciences (Integration Project No.~127).

\appendix




\section{\label{ap:momentum}Momentum equation (\ref{model_eul_a})}

In order to transform (\ref{model_lagr}) into the  Eulerian framework (\ref{model_eul}) we use the following Lagrangian form
\begin{equation}\label{ap_mass}
\frac{{\rm d}\rho}{{\rm d}t}+\rho H_{jk}\frac{\partial v_k}{\partial y_j}=0.
\end{equation}
 of the Eulerian continuity equation $\partial \rho/\partial t+\partial \rho v_k/\partial x_k=0$

We  write now (\ref{model_lagr_a}) as
\begin{equation}\label{ap_mass1}
\rho \frac{{\rm d} v_i}{{\rm d} t}-\rho  F_{k j}H_{l k}\frac{\partial \mathscr{U}_{F_{i j}}}{\partial y_l}=0.
\end{equation}
In this reformulation we have used $ H_{lk}F_{kj}=\delta_{lj}$, where $\delta_{lj}$ is the Kronecker delta. By multiplying  \eqref{ap_mass} by $v_i$ and subsequently adding it to \eqref{ap_mass1} we arrive at
\[v_i\left(\frac{{\rm d} \rho }{{\rm d} t}+\rho  H_{j k}\frac{\partial v_k}{\partial y_j}\right)+\rho \frac{{\rm d} v_i}{{\rm d} t}-\rho  F_{k j}H_{l k}\frac{\partial \mathscr{U}_{F_{ij}}}{\partial y_l}=0.\]
By using
\[H_{lk}\frac{\partial}{\partial y_l}=\dfrac{\partial}{\partial x_k}\]
we get
\[v_i\frac{{\rm d} \rho }{{\rm d} t}+\rho \frac{{\rm d} v_i}{{\rm d} t}+\rho  v_i\frac{\partial v_k}{\partial x_k}-\rho  F_{k j}\frac{\partial \mathscr{U}_{F_{i j}}}{\partial x_k}=0.\]
After adding the identity $\partial \rho F_{kj}/\partial x_k\equiv 0$ multiplied  by  $\mathscr{U}_{F_{ij}}$ to the last equation, we arrive at
\[v_i\frac{{\rm d} \rho }{{\rm d} t}+\rho \frac{{\rm d} v_i}{{\rm d} t}+\rho  v_i\frac{\partial v_k}{\partial x_k}-\frac{\partial \rho  F_{k j} \mathscr{U}_{F_{i j}}}{\partial x_k}=0.\]
Expanding the substantial derivative ${\rm d}/{\rm d}t=\partial/\partial t+v_k(\partial/\partial x_k)$ leads then to
\[\frac{\partial \rho v_i}{\partial t}+\frac{\partial(\rho v_i v_k- \rho  F_{k j} \mathscr{U}_{F_{i j}})}{\partial x_k}=0.\]
Finally, the momentum equation (\ref{model_eul_a}) can be derived from the last
equality by substituting the Cauchy stresses $ \rho  F_{k j} \mathscr{U}_{F_{i j}}$ with the expression \eqref{total_stress_comp}.

\section{\label{ap:rel}Relative velocity equation (\ref{model_eul_d})}

In this appendix we show how to derive the time evolution equation (\ref{model_eul_d})for the relative velocity. By expanding (\ref{model_lagr_d}) we get
\begin{equation}\label{ap_rel_vel}
-\hat{\eta}_j=\frac{{\rm d} \hat{w}_j}{{\rm d} t} + \frac{\partial \mathscr U_c}{\partial y_j} =
\frac{{\rm d} F_{kj}w_j}{{\rm d} t} + \frac{\partial \mathscr U_c}{\partial y_j} =
F_{kj}\frac{{\rm d}w_k}{{\rm d}t}+\left(w_m\frac{\partial v_m}{\partial y_j}+\frac{\partial \mathscr{U}_c}{\partial y_j}\right).
\end{equation}
To arrive at  (\ref{ap_rel_vel}), we used the equality ${\rm d}F_{mj}/{\rm d}t=\partial v_m/\partial y_j$. By multiplying (\ref{ap_rel_vel}) by the inverse strain tensor $\FF^{-1}=\HH =[H_{ij}]$ we obtain
\[-\eta_j=\frac{{\rm d}w_k}{{\rm d}t}+H_{jk}\left(w_m\frac{\partial v_m}{\partial y_j}+\frac{\partial \mathscr{U}_c}{\partial y_j}\right).\]
Substitution of variables
\[H_{jk}\frac{\partial}{\partial y_j}=\dfrac{\partial}{\partial x_k}\]
transforms the previous equation into
\[-\eta_j=\frac{{\rm d}w_k}{{\rm d}t}+w_m\frac{\partial v_m}{\partial x_k}+\frac{\partial \mathscr{E}_c}{\partial x_k}.\]
Here we have also used  $\mathscr{U}_c=\mathscr{E}_c$. By expanding the substantial derivative ${\rm d}/{\rm d}t=\partial/\partial t+v_k(\partial/\partial x_k)$  we arrive at
\[-\eta_j=\frac{\partial  w_k}{\partial t}+v_m\frac{\partial  w_k}{\partial x_m}+w_m\frac{\partial  v_m}{\partial
x_k}+\frac{\partial  \mathscr{E}_c}{\partial x_k}\equiv\frac{\partial  w_k}{\partial t}+v_{m }\frac{\partial  w_{m }}{\partial x_k}+w_{m }\frac{\partial  v_{m }}{\partial x_k}+\frac{\partial
 \mathscr{E}_c}{\partial x_k}+v_{m }\left(\frac{\partial w_k}{\partial x_{m }}-\frac{\partial w_{m }}{\partial x_k}\right),\]
and finally into (\ref{model_eul_d}):
\[\frac{\partial w_k}{\partial t}+\frac{\partial (v_m w_m + \mathscr{E}_c)}{\partial x_k}=-(\varepsilon_{kml}v_m\omega_l+\eta_k).\]

\section{\label{ap:energy}Derivation of the mass and energy conservation}

In this appendix  we prove that the mass conservation (\ref{model_eul_g}) is a consequence of (\ref{model_eul_b}) and that the energy conservation (\ref{eul_en}) is a consequence of (\ref{model_eul}).

First of all, we demonstrate how the continuity equation (\ref{model_eul_g}) follows from Eq.~(\ref{model_eul_b}). It is obvious that if we multiply each equation in (\ref{model_eul_b}) by $\rho_{A_{ik}}$ and sum up the results, the terms with the time derivative $\rho_{A_{ik}}\partial A_{ik}/\partial t$ give $\partial \rho/\partial t$. Subsequently, we recall that $\rho=\rho_0\det\AAA$ and thus $\rho_\AAA=[\rho_{A_{ik}}]=(\rho_0\det\AAA)\AAA^{-\mathsf{T}}=\rho\AAA^{-\mathsf{T}}=\rho\EE^{\mathsf{T}}$.

Now, it is obvious that ${\rm tr}(\rho_\AAA^\mathsf{T}\boldsymbol{\Phi})=\rho_{A_{ik}}\phi_{ik}=0$. It remains to show that
\begin{equation}\label{eq:ap1}
\rho_{A_{ik}}\left(\frac{\partial A_{i k}}{\partial t}+\frac{\partial A_{im} v_m}{\partial x_k}+v_j\left(\frac{\partial A_{ik}}{\partial x_j}-\frac{\partial A_{ij}}{\partial x_k}\right)\right)\equiv \frac{\partial \rho v_k}{\partial x_k}.
\end{equation}
Indeed, the left hand side of this equality is
\begin{eqnarray}
\rho_{A_{ik}}\left(A_{im}\frac{\partial u_m}{\partial x_k}+u_m\frac{\partial A_{ik}}{\partial x_m}\right)=\rho F_{ki}A_{im}\frac{\partial u_m}{\partial x_k}+\rho_{A_{ik}}u_m\frac{\partial A_{ik}}{\partial x_m}=\rho \frac{\partial u_k}{\partial x_k}+\rho_{A_{ik}}u_m\frac{\partial A_{ik}}{\partial x_m}=\nonumber \\
\rho \frac{\partial u_k}{\partial x_k}+u_m\frac{\partial \rho}{\partial x_m}=\frac{\partial \rho v_k}{\partial x_k}.\nonumber
\end{eqnarray}

Now, we are ready to derive (\ref{eul_en}) from (\ref{model_eul}). We demonstrate this in the case of nonlinear elasticity (it is easy to generalized the proof to the entire system (\ref{model_eul}) by analogy):
\begin{eqnarray}
\displaystyle\frac{\partial \rho v_i}{\partial t}+\frac{\partial (\rho v_i v_k+\rho A_{mi}\mathscr{E}_{A_{mk}})}{\partial x_k}=0,&\label{eq:ap_a}\\[2mm]
\displaystyle\frac{\partial A_{i k}}{\partial t}+\frac{\partial A_{im} v_m}{\partial x_k}+v_j\left(\frac{\partial A_{ik}}{\partial x_j}-\frac{\partial A_{ij}}{\partial x_k}\right)=0,&\label{eq:ap_b}\\[2mm]
\displaystyle\frac{\partial \rho \mathscr{ E}}{\partial t}+\frac{\partial \left(v_k\rho \mathscr{E}+\rho v_n  A_{mn}\mathscr{E}_{A_{mk}}\right) }{\partial x_k}=0.&
\end{eqnarray}

Here, the total energy $\rho\mathscr{E}=\rho(U+v_mv_m/2)=\rho U(\boldsymbol{A})+(\rho v_m)(\rho v_m)/(2\rho)$ is a function of the independent variables $\rho \vv$, $\boldsymbol{A}$. As above, $\rho=\rho_0\det\boldsymbol{A}$. Consequently, $(\rho\mathscr{E})_{\rho v_i}=v_i$ and $(\rho\mathscr{E})_{A_{ij}}=\rho_{A_{ij}}(U-v_m v_m/2)+\rho U_{A_{ij}}$. It is obvious that
\[v_i\frac{\partial \rho v_i}{\partial t}+(\rho_{A_{ij}}(U-v_m v_m/2)+\rho U_{A_{ij}})\frac{\partial A_{i k}}{\partial t}
\equiv (\rho\mathscr{E})_{\rho v_i}\frac{\partial \rho v_i}{\partial t}+(\rho\mathscr{E})_{A_{ij}}\frac{\partial A_{i k}}{\partial t}\equiv \frac{\partial \rho \mathscr{ E}}{\partial t}.\]
Therefore, it remains to show that
\begin{eqnarray}\label{ap:ABC}\frac{\partial \left(v_k\rho \mathscr{E}+\rho v_n  A_{mn}\mathscr{E}_{A_{mk}}\right) }{\partial x_k}-(\rho\mathscr{E})_{\rho v_i}\frac{\partial (\rho v_i v_k+\rho A_{mi}\mathscr{E}_{A_{mk}})}{\partial x_k}-\qquad\qquad\qquad\qquad\qquad\nonumber\\[2mm]
(\rho\mathscr{E})_{A_{ij}}\left[\frac{\partial A_{im} v_m}{\partial x_k}+v_j\left(\frac{\partial A_{ik}}{\partial x_j}-\frac{\partial A_{ij}}{\partial x_k}\right)\right] \equiv A+B+C\equiv 0.
\end{eqnarray}
Now, we expand each terms on the left hand side:
\begin{eqnarray}
& A=\rho  v_kU_{A_{i j}}\frac{\partial  A_{i j}}{\partial x_k}+\rho  v_kv_m\frac{\partial  v_m}{\partial x_k}+\left(U+\frac{v_m v_m}{2}\right)\frac{\partial
 \rho  v_k}{\partial x_k}+\rho  U_{A_{m  i}}A_{m  k}\frac{\partial  v_i}{\partial x_k}+v_i\frac{\partial }{\partial x_k}\left(\rho  U_{A_{m
 i}}A_{m  k}\right),\label{ap:A}\\[2mm]
& B=-v_i v_i\frac{\partial  \rho  v_k}{\partial x_k}-\rho  v_i v_k\frac{\partial  v_i}{\partial x_k}-v_i\frac{\partial \left(\rho  A_{m  k}U_{A_{m
  i}}\right)}{\partial x_k},\label{ap:B}\\[2mm]
& C=-\rho  A_{i m }U_{A_{i k}}\frac{\partial  v_{m }}{\partial x_k}-\rho  v_{m }U_{A_{i k}}\frac{\partial A_{i m }}{\partial x_k}-\rho
  U_{A_{i k}}v_{m }\frac{\partial A_{i k}}{\partial x_{m }}+\rho  U_{A_{i k}}v_{m }\frac{\partial A_{i m }}{\partial x_k}-\left(U-\frac{v_mv_m}{2}\right)
 \frac{\partial  \rho  v_k}{\partial x_k}.\label{ap:C}
\end{eqnarray}
In the expansion of $C$ we have used (\ref{eq:ap1}). It is easy to see that the first term in (\ref{ap:A}) is canceled by the second, third and fourth terms in (\ref{ap:C}). The second and third terms in (\ref{ap:A}) are canceled by the first and second terms in (\ref{ap:B}) and by the last term in (\ref{ap:C}). The fourth term in (\ref{ap:A}) is canceled by the first term in (\ref{ap:C}). Finally, the last term in (\ref{ap:A}) is canceled by the last term in (\ref{ap:B}). This ends the proof of (\ref{ap:ABC}).

\section{\label{ap:matrix}Entries of the matrix (\ref{matrixA})}

In this appendix we give   formulas  for entries  of the matrix (\ref{matrixA}). The following notation is used below: $\mathscr{U}_{\FF\FF}=[\mathscr{U}_{F_{ij}F_{mn}}]$, $U_{\FF\FF}=[U_{F_{ij}F_{mn}}]$ are $9\times9$-matrices, $\mathscr{U}_{\FF c}=[\mathscr{U}_{F_{ij}c}]$, $U_{\FF c}=[U_{F_{ij}c}]$ are nine-dimensional vectors, $\mathscr{U}_{\FF{\hat{\ww}}}=[\mathscr{U}_{F_{ij}\hat{w}_m}]$ is $9\times3$-matrix, $\mathscr{U}_{{\hat{\ww}}{\hat{\ww}}}=[\mathscr{U}_{\hat{w}_i\hat{w}_j}]$, $W_{\ww\ww}=[W_{w_i w_j}]$ are $3\times3$-matrix, $\mathscr{U}_{{\hat{\ww}}c}=[\mathscr{U}_{\hat{w}_ic}]$, $W_{\ww c}=[W_{{w}_ic}]$ are three-dimensional vectors, symbol $\otimes$ denotes the Kronecker product and $\HH=\FF^{-1}$. In all formulas below, it is implied that the function $W$ is a function of $c$ and $\ww$ as in the Eulerian framework.  The entries of the matrix (\ref{matrixA}) are:
\[
\mathscr{U}_{\FF\FF}=U_{\FF\FF}+\HH^\mathsf{T}\otimes(\ww (\HH W_\ww )^\mathsf{T})+((\HH W_\ww )\ww ^\mathsf{T})\otimes\HH ^\mathsf{T}
+(\HH W_{\ww\ww}\HH ^\mathsf{T})^\mathsf{T}\otimes({\ww\ww}^\mathsf{T}),
\]
\[
\mathscr{U}_{\FF{\hat{\ww}}}=-(\HH W_{\ww\ww}\HH ^\mathsf{T})\otimes \ww - (\HH W_\ww )\otimes\HH ^\mathsf{T},
\]
\[
\mathscr{U}_{\FF c}=U_{\FF c}-(\HH W_{\ww c})\otimes\ww ,\ \ \ \ \ \mathscr{U}_{{\hat{\ww}}{\hat{\ww}}}=\HH W_{\ww\ww}\HH ^\mathsf{T},\ \ \ \ \ \ \mathscr{U}_{{\hat{\ww}}c}=\HH W_{\ww c}.\]

In case of irreversible deformation (i.e. $\FF=\EE\PP$, $\PP\neq\II$), the matrix $U_{\FF\FF}$ is \cite{pesh2010}
\[U_{\FF\FF}=\mathcal{P}U_{\EE\EE}\mathcal{P}^\mathsf{T},\]
where $\mathcal{P}=\PP^{-1}\otimes\II$.

\bibliographystyle{vancouver}
\bibliography{ref_cmt}

\end{document}